\documentclass[prd,superscriptaddress,a4paper]{revtex4}

\usepackage{graphicx}
\usepackage{hyperref}
\usepackage{bm}
\usepackage{amsmath}
\usepackage{amssymb}
\usepackage{color}

\newcommand{\hMpc}{\ h^{-1}\text{Mpc}}

\newcommand{\ihMpc}{\ h\text{Mpc}^{-1}}

\newcommand{\hMs}{\ h^{-1} M_\odot}
\newcommand{\eh}[1]{\exp{\left[#1\right]}}
\newcommand{\tim}[1]{\times 10^{#1}}

\newcommand{\derd}{\,\mathrm{d}} 

\newcommand{\be}{\begin{equation}}
\newcommand{\ee}{\end{equation}}
\newcommand{\la}{\left\langle}
\newcommand{\ra}{\right\rangle}
\newcommand{\derivd}{\text{d}}
\renewcommand{\vec}{\bm}

\newcommand{\dkpc}{\frac{\derivd^3k'}{(2\pi)^3}}

\newcommand{\ii}{{\text{i}\,}}

\begin{document}
\title{Evidence for Quadratic Tidal Tensor Bias from the Halo Bispectrum}
\author{Tobias Baldauf}
\email{baldauf@physik.uzh.ch}
\affiliation{Institute for Theoretical Physics, University of Zurich, Zurich, Switzerland}
\author{Uro\v{s} Seljak}
\affiliation{Institute for Theoretical Physics, University of Zurich, Zurich, Switzerland}
\affiliation{Lawrence Berkeley National Laboratory, Berkeley, CA, USA}
\affiliation{Physics Dept.\ and Astronomy Dept., University of California, Berkeley, CA, USA}
\affiliation{Institute for the Early Universe, EWHA Womans University, Seoul, South Korea}
\author{Vincent Desjacques}
\affiliation{Department of Theoretical Physics and Center for Astroparticle Physics, University of Geneva, Geneva, Switzerland}
\author{Patrick McDonald}
\affiliation{Lawrence Berkeley National Laboratory, Berkeley, CA, USA}
\affiliation{Physics Dept., Brookhaven National Laboratory, Upton, NY, USA}
\date{\today}
\begin{abstract}
The relation between the clustering properties of luminous matter in the form of galaxies and the underlying dark matter distribution is of fundamental importance for the interpretation of ongoing and upcoming galaxy surveys. The so called local bias model, where galaxy density is a function of local matter density, is frequently discussed as a means to infer the matter power spectrum or correlation function from the measured galaxy correlation. However, gravitational evolution generates a term quadratic in the tidal tensor and thus non-local in the Eulerian density field, even if this term is absent in the initial conditions (Lagrangian space). Because the term is quadratic, it contributes as a loop correction to the power spectrum, so the standard linear bias picture still applies on very large scales, however, it contributes at leading order to the bispectrum for which it is significant on all scales. Such a term could also be present in Lagrangian space if halo formation were influenced by the tidal field. We measure the corresponding coupling strengths from the matter-matter-halo bispectrum in numerical simulations and find a non-vanishing coefficient for the tidal tensor term. We find no scale dependence of the inferred bias parameters up to $k \sim 0.1\ihMpc$ and that the tidal effect is increasing with halo mass. While the local Lagrangian bias picture is a better description of our results than the local Eulerian bias picture, our results suggest that there might be a tidal tensor bias already in the initial conditions. We also find that the coefficients of the quadratic density term deviate quite strongly from the theoretical predictions based on the spherical collapse model and a universal mass function. Both quadratic density and tidal tensor bias terms must be included in the modeling of galaxy clustering of current and future surveys if one wants to achieve the high precision cosmology promise of these datasets. 

\end{abstract}
\maketitle

\section{Introduction}
Large Scale Structure (LSS), the large-scale distribution of matter in the 
Universe, contains a wealth of information about the history and composition of the Universe as well as fundamental physics. For instance, LSS has the potential to constrain neutrino masses or modifications of gravity, which however requires percent level accuracy for the theory and observations. Besides gravitational lensing, which is sensitive to the total matter distribution, the positions of galaxies are the main observable and tool to infer the underlying matter distribution. They have the advantage of higher statistical power relative to weak lensing surveys. Ongoing and upcoming LSS surveys such as BOSS, BigBOSS, EUCLID, DES will provide an unprecedented quality of galaxy clustering data, which needs to be properly analyzed. 

A crucial step in the analysis of galaxy surveys is to connect the distribution of the tracer to the underlying distribution of matter. The first step in this logical chain is the realization that galaxies form preferentially in the potential wells of collapsed dark matter halos, where the hot gas can cool sufficiently fast \cite{1977MNRAS.179..541R,1978MNRAS.183..341W}. This leads to the question of how the clustering properties of dark matter halos relate to the clustering of matter in general. The answer to this question is usually phrased in terms of a relation between the overdensities in these two fields and is dubbed a halo biasing scheme. 
With the ever increasing computing power it is in principle possible to generate templates for the survey analysis for standard $\Lambda$CDM and even modified gravity models using $N$-body simulations. This approach becomes very expensive when it is to be used in Markov-Chain-Monte-Carlo parameter inference methods and does not provide insight into the underlying clustering properties. We thus consider it important to understand the properties of halo clustering by testing theoretical prescriptions on simulations with the final goal of devising \emph{analytical} and thus easily evaluable models for the survey analysis.

The so called local biasing model \cite{Kaiser:1984sw,Fry:1992vr}, where galaxy and halo density is a function of local matter density, has been the most popular model used in previous work.  In the simplest version one adds another contribution that scales quadratically with density, the quadratic density bias term. Recent work has argued, based on symmetry and analyticity arguments, that there are additional terms not included in the local bias model that appear in the power spectrum and are formally at the same order as that of quadratic bias \cite{McDonald:2009dh}. One of these terms is quadratic in density but non-local and can be written as the square of the tidal tensor. The first goal of this paper is to provide additional theoretical motivation for inclusion of this term in the analysis of galaxy clustering. 

The working horses in LSS analysis are the two-point functions, the correlation function and the power spectrum. Purely Gaussian, linear fields are completely characterized by their two point function. However, non-linear phenomena in galaxy and halo formation as well as non-linear gravitational clustering can generate the full hierarchy of $n$-point functions.  These higher order statistics might be difficult to measure in the sky due to non-trivial survey windows and redshift space distortions, but they can be easily extracted from $N$-body simulations. The simplest statistic beyond the power spectrum is the bispectrum
\be
\la \delta(\vec k_1)\delta(\vec k_2)\delta(\vec k_3)\ra= B\left(\vec k_1,\vec k_2,\vec k_3\right)\ (2\pi)^3\delta^\text{(D)}(\vec k_1+\vec k_2+\vec k_3),
\ee
which is well suited for the study of next-to-leading order effects in cosmic density fields \cite{1994ApJ...425..392F,1997MNRAS.290..651M,1999ApJ...521....1B,2005MNRAS.361..824G,2011MNRAS.415..383M}. While these contribute only loop terms to the power spectrum they are the leading order terms for the bispectrum, which vanishes for purely linear Gaussian fields.

The second aim of this study is to probe the halo bispectrum for Gaussian initial conditions in the low-$k$ regime, in order to extract the two terms that lead to a quadratic coupling between the number density of collapsed objects and the long wavelength matter fluctuations, quadratic density bias and quadratic tidal tensor bias. We present a study of their scale and mass dependence using $N$-body simulations and we compare the numerical results to the theoretical expectations based on simple halo bias models. 

This paper is organized as follows. In \S \ref{sec:biasmodel} we review the standard formulation of the bias model and discuss possible extensions. \S \ref{sec:simulations} describes the simulations, the bispectrum measurement and data reduction as well as the parameter estimation. The results are presented in \S \ref{sec:results}. Finally, in \S \ref{sec:discussion} we discuss our findings and their implications as well as possible directions for future investigation.
\section{The bias model: local and non-local forms}\label{sec:biasmodel}
\subsection{Standard Formulation: Local Bias}
The formation of galaxies and their host dark matter halos is a complicated highly non-perturbative process. It is, however, reasonable to assume that certain properties of collapsed objects, for instance their number density, are related to the coarse grained underlying matter density field in the same region of space. Neglecting complications arising from gas physics, the number density of collapsed objects can be written as a functional of the underlying matter density perturbation \cite{Fry:1992vr}
\be
\delta_\text{h}\left(\vec x,\eta \right)=\mathcal{F}\left[\delta(\vec x',\eta)\right].
\ee
On large scales, this functional is commonly approximated by a \emph{local} and
\emph{linear} bias model 
$\delta_\text{h}\left(\vec x,\eta\right)=b_1 \delta\left(\vec x,\eta\right)$ 
(plus generally a noise term which we will avoid in this paper by only looking
at cross-correlations with mass). 
The next step is to give up on linearity and to introduce the second order \emph{local} bias model $\delta_\text{h}\left(\vec x,\eta\right)=b_1 \delta\left(\vec x,\eta\right)+b_2 \delta^2\left(\vec x,\eta\right)$. This model has been well studied in the literature in combination with Standard Perturbation Theory (SPT) and leads to non-trivial renormalizations of the leading order bias parameter \cite{McDonald:2006mx}.
Measurements of the quadratic bias parameters of this model in the bispectrum 
have lead to contradictory results, which raises doubts about the completeness 
of the model \cite{Pollack:2011xp} (see also \cite{2011MNRAS.415..829R}).

\subsection{Tidal Terms}
The power series expansion of the functional presented above is certainly overly simplified and one should consider whether other terms could influence the number density of collapsed objects. 
As proposed in \cite{McDonald:2009dh}, the environmental dependence of halo formation could lead to a dependence on the tidal field, as quantified by the tidal tensor
\be
s_{ij}(\vec x,\eta)=\partial_i \partial_j
\Phi(\vec x,\eta)-\frac{1}{3}\delta^\text{(K)}_{ij}\delta(\vec x,\eta).
\ee
Note that we absorbed the constants in the Poisson equation into the gravitational potential $\vec \nabla^2 \Phi(\vec x,\eta)=\delta(\vec x,\eta)$ and subtract out the trace from the tidal tensor because it is degenerate with the density field.
The corresponding expression in Fourier space is given by
\be
s_{ij}(\vec k,\eta)=\left(\frac{k_i
k_j}{k^2}-\frac{1}{3}\delta_{ij}^\text{(K)}\right)\delta(\vec k,\eta).
\ee
The halo overdensity is a scalar quantity and can thus only depend on scalars. The simplest scalar that can be constructed from the tidal tensor is given by $s^2(\vec x)=s_{ij}(\vec x)s_{ij}(\vec x)$ which in Fourier space is expressed by the convolution
\be
s^2(\vec k,\eta)=\int \dkpc S_2(\vec k',\vec k-\vec k')\delta(\vec k',\eta)\delta(\vec k - \vec
k',\eta)\ ,
\ee
where we have implicitly defined the kernel
\be
S_2(\vec q_1,\vec q_2)=\frac{\left(\vec q_1 \cdot \vec q_2\right)^2}{\vec{q_1}^2
\vec q_2^2}-\frac{1}{3}\ .
\ee

Following \cite{McDonald:2009dh}, the halo density field up to second order 
can be written as
\be
\delta_\text{h}(\vec x,\eta)=b_1 \delta(\vec x,\eta) +b_2 \bigl[\delta^2(\vec x,\eta)-\la\delta^2(\vec x,\eta)\ra\bigr]+b_{s^2}\bigl[s^2(\vec x,\eta)-\la s^2(\vec x,\eta)\ra\bigr],
\label{eq:densitys2}
\ee
where we absorbed prefactors of $1/2$ into the bias parameters. 
We truncated the series at second order, since higher order terms influence the bispectrum only through loop corrections, which are believed to be subdominant on large scales. As can be easily verified from the above definitions one has $\bigl\langle s^2(\vec x,\eta)\bigr\rangle=2/3\ \bigl\langle {}^{(1)}\delta^2(\vec x,\eta)\bigr\rangle$.
\subsection{Lagrangian Bias}
In the usual Lagrangian bias picture, the galaxy formation sites are identified
in the primordial density field, and it is assumed that the primordial halo 
density field at initial time $\eta_\text{i}$ can be written as a power series 
in the primordial matter fluctuations. For calculational convenience, the 
expansion can be rewritten in terms of the linearly extrapolated density field 
$\delta(\vec q,\eta)=D(\eta)/D(\eta_i) \delta(\vec q)$
\be
\delta_\text{h}(\vec q)=
\sum_l \frac{b_l^\text{(L)}(\eta_\text{i})}{l!}\left[\delta^l(\vec q)-
\left<\delta^l(\vec q)\right>\right]=
\sum_l \frac{b_l^\text{(L)}(\eta)}{l!}\left[\delta^l(\vec q,\eta)-
\left<\delta^l(\vec q,\eta)\right>\right],
\label{eq:lagrangianbias}
\ee
where  $b_l^\text{(L)}(\eta)=\bigl(D(\eta_i)/D(\eta)\bigr)^l\, b_l^\text{(L)}(\eta_i)$ and $\vec q$ is the Lagrangian position. Here, we introduced the conformal time $a \derd \eta=\derd t$ and the linear growth factor $D(\eta)$, normalized to unity at present time.
We start the sum from $l=1$ because the $l=0$ term vanishes by requiring that the halo field has a vanishing mean. 
In contrast to this, the Eulerian bias model expands the halo density field at a certain point in time in the non-linear matter density field at the same time. Motivation for a Lagrangian nature of halo bias comes from the peak model \cite{Bardeen:1985tr,1989MNRAS.238..293L,Desjacques:2010css,Elia:2011ds}, where the peaks of the primordial density field are associated with the formation sites of protohaloes.
\par
It now remains to connect the Lagrangian density fields to the observable Eulerian ones. The continuity equation for halos requires
\be
\left[1+\delta_\text{h}(\vec x,\eta)\right]\derd^3 x=\left[1+\delta_\text{h}(\vec q)\right]\derd^3 q,
\label{eq:halocontinuity}
\ee
where
\be
\left[1+\delta(\vec x,\eta)\right] \derd^3 x=\derd^3 q
\ee
is the continuity equation for the underlying dark matter field.
Note that the Lagrangian density field we work with is always the linear 
density field, because this was the definition of the bias expansion in 
Eq.~\eqref{eq:lagrangianbias} -- more general expansions are possible in 
principle. 

\par
The Lagrangian and Eulerian positions are related by $\vec x(\vec q,\eta)=\vec q +\vec \Psi (\vec q,\eta)$, thus up to third order in the density field we have 
\be
\delta(\vec q)=\delta(\vec x,\eta)-\Psi_i(\vec q,\eta) \partial_i \delta(\vec q) +\frac{1}{2}\Psi_i(\vec q,\eta) \Psi_j(\vec q,\eta) \partial_i \partial_j \delta(\vec q).
\ee
In contrast to Eulerian perturbation theory, where the density is the central quantity, Lagrangian Perturbation Theory (LPT) has the displacement field $\vec \Psi$ as the central dynamic quantity and the density fields are only derived quantities (for the basics of LPT and its relation to SPT see Appendix \ref{app:lpt}). For simplicity, we will focus on the matter-only Einstein-de Sitter Universe for the theoretical calculations throughout this paper.
Using the Lagrangian bias expansion \eqref{eq:lagrangianbias} in \eqref{eq:halocontinuity} and expressing the Lagrangian position field in terms of the Eulerian position, we have (see also \cite{Catelan:2000vn})
\begin{align}
\delta_\text{h}(\vec x,\eta)=&\left(1+b_1^\text{(L)}(\eta)\right){}^{(1)}
\delta(\vec x,\eta)\nonumber\\
+& b_1^\text{(L)}(\eta) \delta^2(\vec x,\eta)
+\frac{1}{2}b_2^\text{(L)}(\eta)\Bigl[{}^{(1)}
\delta^2(\vec x,\eta)-\bigl\langle{}^{(1)} \delta^2(\vec x,\eta)\bigr\rangle\Bigr]
-b_1^\text{(L)}(\eta)\vec \Psi(\vec q,\eta) \cdot \vec 
\nabla \delta(\vec x,\eta)+{}^{(2)}
\delta(\vec x,\eta).
\end{align}
Reorganizing the terms in order to have the first order bias multiply the full second order matter density field, we obtain
\begin{align}
\delta_\text{h}(\vec x,\eta)=&\left(1+b_1^\text{(L)}(\eta)\right)
\left(^{(1)} \delta(\vec x,\eta)+{}^{(2)}\delta(\vec x,\eta)\right)\nonumber\\
+&\left(\frac{4}{21}b_1^\text{(L)}+\frac12 b_2^\text{(L)}(\eta)\right) 
\Bigl[{}^{(1)}\delta^2(\vec x,\eta)-\bigl\langle{}^{(1)}\delta^2(\vec x,\eta)\bigr\rangle \Bigr]-\frac{2}{7}b_1^\text{(L)}(\eta) \Bigl[s^2(\vec x,\eta)-\bigl\langle s^2(\vec x,\eta)\bigr\rangle\Bigr],
\label{eq:lagbias}
\end{align}
where we used that the second order mass density (in SPT and LPT) can be written in configuration space as (see e.g.~\cite{Bouchet:1992})
\be
{}^{(2)}\delta(\vec x,\eta)=\frac{17}{21}{}^{(1)}\delta^2(\vec x,\eta)-\vec \Psi(\vec x,\eta) \cdot \vec \nabla \delta(\vec x,\eta) +\frac{2}{7} s^2(\vec x,\eta).
\label{eq:secondorderdensity}
\ee
We see that the functional form of the above result agrees with Eq.~\eqref{eq:densitys2}, but complements it with a dynamical perspective. In particular, we see that even in the absence of a tidal tensor bias in the initial conditions, such a term will be generated by subsequent gravitational evolution.
While our result in Eq.~\eqref{eq:lagbias} is consistent with the relations presented in \cite{Catelan:2000vn}, we have reorganized terms in order to have the first order bias multiplying the second order density field Eq.~\eqref{eq:secondorderdensity} and to make the formal equivalence with the phenomenological picture presented above more obvious. Furthermore, there are no $b_0^\text{(L)}$ factors, because we fixed the original Taylor series for $\delta_\text{h}(\vec q)$  to have vanishing mean.

\subsection{Coevolution of Halos and Dark Matter}
In this subsection we will consider an (at first glance) Eulerian approach to 
halo clustering and consider the coevolution of the coupled halo-dark matter 
fluid (For a similar approach in combination with resummed perturbation theory 
see \cite{Elia:2010en}).
Assuming vanishing velocity bias $\vec v_\text{h}=\vec v$ and 
\emph{conservation of halo number}, we can write down a coupled system of 
differential equations for the matter and halo fluid 
\footnote{A similar approach was presented by R.~Scoccimarro at the PTchat 
workshop in Saclay.}, namely the continuity equation for halos, the continuity 
equation for the matter and the combined Euler and Poisson equations for matter
\begin{align}
\delta_\text{h}'(\vec k,\eta)+\theta(\vec k,\eta)=&-\int \dkpc \alpha(\vec k',\vec k-\vec k') \theta(\vec k',\eta)\delta_\text{h}(\vec k-\vec k',\eta)\label{eq:halocontinuityk}\\
\delta'(\vec k,\eta)+\theta(\vec k,\eta)=&-\int \dkpc \alpha(\vec k',\vec k-\vec k') \theta(\vec k',\eta)\delta(\vec k-\vec k',\eta)\\
\theta'(\vec k,\eta)+\mathcal{H}(\eta)\theta(\vec k,\eta)+\frac32 \Omega_\text{m}(\eta)\mathcal{H}^2(\eta)\delta(\vec k,\eta)=&-\int \dkpc \beta(\vec k',\vec k-\vec k')\theta(\vec k',\eta)\theta(\vec k-\vec k',\eta)
\end{align}
Here we introduced the velocity divergence $\theta(\vec x)=\vec \nabla \cdot \vec v(\vec x)$. The matter equations at first order are solved by ${}^{(1)}\theta(\vec k,\eta)=-\mathcal{H}(\eta){}^{(1)}\delta(\vec k,\eta)$ and ${}^{(1)}\delta(\vec k,\eta)=D(\eta) {}^{(1)}\delta_0(\vec k)$, where ${}^{(1)}\delta_0(\vec k)$ is the present day linear overdenisty. The second order solution for the matter yields
\begin{align}
{}^{(2)}\delta(\vec k,\eta)=&\int \dkpc F_2(\vec k',\vec k-\vec k'){}^{(1)}\delta(\vec
k',\eta){}^{(1)}\delta(\vec k-\vec k',\eta)\\
{}^{(2)}\theta(\vec k,\eta)=&-\mathcal{H}(\eta)\int \dkpc G_2(\vec k',\vec k-\vec k'){}^{(1)}\delta(\vec
k',\eta){}^{(1)}\delta(\vec k-\vec k',\eta)
\end{align}
where the second order SPT mode coupling kernels are defined as \cite{Bernardeau:2001qr}
\begin{align}
F_{2}(\vec q_1,\vec q_2)=&\frac{5}{7}\alpha(\vec q_1,\vec q_2)+\frac{2}{7}\beta(\vec q_1,\vec q_2)\\
G_{2}(\vec q_1,\vec q_2)=&\frac{3}{7}\alpha(\vec q_1,\vec q_2)+\frac{4}{7}\beta(\vec q_1,\vec q_2)
\end{align}
with
\begin{align}
\alpha(\vec q_1,\vec q_2)=\frac{\left(\vec q_1+\vec q_2\right)\cdot \vec q_1}{
q_1^2} && \beta(\vec q_1,\vec q_2)=\frac{1}{2}\left(\vec q_1 +\vec
q_2\right)^2\frac{\vec q_1 \cdot \vec q_2}{q_1^2 q_2^2}
\end{align}
\par
The first order equation for the halos can be solved using the local bias ansatz ${}^{(1)}\delta_\text{h}(\vec k,\eta)=b_1^\text{(E)}(\eta){}^{(1)}\delta(\vec k,\eta)$, which then gives the time evolution of the first order bias as
\be
\frac{b_1^\text{(E)}(\eta)-1}{b_{1}^\text{(E)}(\eta_\text{i})-1}=\frac{D(\eta_\text{i})}{D(\eta)}.
\ee
This relation known as debiasing, i.e., at late times the bias converges to unity and halo and matter density field agree \cite{Fry:1996fg}.
\par
Solving Eq.~\eqref{eq:halocontinuityk} at second order using the second order matter solutions, we obtain
\begin{align}
^{(2)}\delta_\text{h}(\vec k,\eta)&={}^{(2)}\delta_\text{h}(\vec k,\eta_\text{i})+b_1^\text{(E)}(\eta) \int  \dkpc F_2(\vec k',\vec k-\vec k')\delta(\vec k',\eta)\delta(\vec k-\vec k',\eta)\nonumber\\
&+\frac{4}{21}\left(b_1^\text{(E)}(\eta)-1\right) \int\dkpc \delta(\vec k',\eta)\delta(\vec k-\vec k',\eta)\\
&-\frac{2}{7}\left(b_1^\text{(E)}(\eta)-1\right)\int\dkpc S_2(\vec k',\vec k -\vec k')\delta(\vec k',\eta)\delta(\vec k-\vec k',\eta)\nonumber ,
\end{align}
where we assumed $D(\eta_i)\ll D(\eta)$.
Here, we have isolated the part proportional to the second order matter field in the first line.
Thus, the dynamical evolution naturally introduces a $\delta^2(\vec x)$ and a $s^2(\vec x)$ term, even in the absence thereof at some initial time $\eta_\text{i}$. 
Translating back to real space, we see that this has the same functional form 
as Eq.~\eqref{eq:densitys2} and agrees with the Lagrangian bias picture. The 
equivalence is even more obvious if we have 
${}^{(2)}\delta_\text{h}(\vec x,\eta_\text{i})=b_2^\text{(L)}(\eta_\text{i}) 
{}^{(1)}\delta^2(\vec x,\eta_\text{i})/2=b_2^\text{(L)}(\eta){}^{(1)}
\delta^2(\vec x,\eta)/2$ in correspondence to the Lagrangian bias picture 
discussed above (see next subsection for a relation between the parameters of 
the models discussed here).
We note in retrospect that, while the calculation took an Eulerian form, the
specification that galaxies formed at an early time 
tracing the initial density field  (or at least the location where they will 
form is determined this way), and then just follow 
gravity is actually the same one made in the Lagrangian calculation.

\subsection{Relation between Eulerian and Lagrangian bias parameters} \label{sec:biasrelation}
As we have seen above, the Lagrangian bias model, the coevolution model and the educated guess of \cite{McDonald:2009dh} all lead to the same functional form for the halo density field if one identifies the parameters of the model as
\begin{align}
b_1=b_1^\text{(E)}(\eta)=&1+b_1^\text{(L)}(\eta)\\
b_2=\frac{1}{2}b_2^\text{(E)}(\eta)=&\frac{4}{21}b_1^\text{(L)}(\eta)+ \frac{1}{2}b_2^\text{(L)}(\eta)~.
\label{eq:quadraticbias}
\end{align}
This identification agrees with the one of the 
spherical collapse picture \cite{Scoccimarro:2000gm}. 
Note however that our results are \emph{not} relying on the spherical collapse dynamics. For the prefactor of the tidal field scalar we have
\be
b_{s^2}=-\frac{2}{7}\left(b_1^\text{(E)}(\eta)-1\right)=-\frac{2}{7}b_1^\text{(L)}(\eta).
\ee
In this model there is no tidal field bias in Lagrangian space, consistent with the spherical collapse model, 
although in the ellipsoidal collapse model \cite{1979ApJ...231....1W,1996ApJS..103....1B,ShethMoTormen,Desjacques:2007zg} such a term would be allowed. 

When comparing our measurements to theoretical bias models, we consider the bias derived from the Sheth-Tormen \cite{Sheth:1999mn} mass function with parameters $p=0.15$ and $q=0.75$ that were shown to be in good agreement with first order bias in $N$-body simulations \cite{Seljak:2004ni}, although we checked that the predictions do not differ much from those using the original values in \cite{Sheth:1999mn}. The Lagrangian bias parameters are then given by the first and second derivative of the mass function $n(M)$ with respect to a long wavelength background fluctuation $\delta_\text{l}$. For universal mass functions these can be rewritten as derivatives with respect to the peak height $\nu(M,\eta)=\delta_\text{c}^2/\sigma^2(M,\eta)$
\begin{align}
b_{1}^\text{(L)}=&\frac{1}{\bar{n}}\frac{\partial n}{\partial \delta_\text{l}}=-\frac{1}{\bar
n}\frac{2\nu}{
\delta_\text{c} }
\frac{\partial n}{\partial \nu}\\
b_{2}^\text{(L)}=&\frac{1}{\bar{n}}\frac{\partial^2 n}{\partial \delta_\text{l}^2}=\frac{4}{\bar
n}
\frac{\nu^2}{\delta_c^2}\frac{\partial^2 n}{\partial \nu^2}+\frac{2}{\bar{n}}\frac{\nu}{
\delta_\text{c}^2}\frac{\partial n}{\partial \nu},
\end{align}
where $\delta_\text{c}=1.686$ is the critical density for spherical collapse. The derivatives of the Sheth-Tormen mass function read
\begin{align}
\frac{1}{n}\frac{\partial n}{\partial
\nu}=&-\frac{q\nu-1}{2\nu}-\frac{p}{\nu\left(1+(q\nu)^p\right)}\\
\frac{1}{n}\frac{\partial^2 n}{\partial
\nu^2}=&\frac{p^2+\nu p
q}{\nu^2\left(1+(q\nu)^p\right)}+\frac{(q\nu)^2-2q\nu-1}{4\nu^2}.
\end{align}
The mass dependence of the bias function is presented in Figure \ref{fig:massdep} and discussed in \S \ref{sec:results}.
\subsection{Bispectra}
The leading order contribution to the bispectrum arises from quadratic terms in the fields. Higher order couplings enter only through loop corrections, which gain importance for high $k$. This is equivalent to the situation in the power spectrum, where linear terms in the field are dominant on large scales and loop corrections from quadratic and higher order terms gain importance for high $k$. 

The tree-level matter bispectrum in SPT is given by
\be
B_\text{mmm}(\vec k_1,\vec k_2,\vec k_3)
=2P(k_1)P(k_2)F_2(\vec k_1,\vec k_2)+2\, \text{cyc.} \ ,
\label{eq:bmmmtheo}
\ee
where cyc.\ symbolizes the two cyclic permutations of the $k$-vectors in the power spectrum and mode coupling kernel.
From an observational point of view the halo auto bispectrum $B_\text{hhh}$ is probably the most appealing statistic. Unfortunately it is suffering from shotnoise, which might deviate from its fiducial Poisson form $1/\bar{n}$  \cite{Hamaus:2010im}.
Besides the halo auto power spectrum, there are two halo-matter cross bispectra $B_\text{mhh}$ and $B_\text{mmh}$, where either one or two matter fields are correlated with two or one halo fields, respectively. One further needs to state whether the cross bispectra are symmetrized over the $\vec k$ modes or not. 

Our focus is not on observability but on understanding the clustering properties of dark matter halos in $N$-body simulations and devising a theoretical framework that can later be used to analyze real data. Thus, we will consider the un-symmetrized matter-matter-halo cross bispectrum defined as
\be
B_\text{mmh}^\text{(unsym)}(\vec k_1,\vec k_2,\vec k_3)\ (2\pi)^3\delta^\text{(D)}(\vec k_1+\vec k_2+\vec k_3)=\la \delta(\vec k_1)\delta(\vec k_2)\delta_\text{h}(\vec k_3)\ra,
\ee
where the halo density field is always on the $\vec{k}_3$-mode.   
This particular configuration might not be the one with the highest signal to noise ratio but it has a couple of quite useful properties for our study: a) The cross-bispectrum does not suffer from a spurious shotnoise contamination and is thus a clean probe of the clustering of halos b) the functional simplicity of the second order bias contributions in terms of $k_1$, $k_2$ and the enclosed angle $\mu$ where the $b_2$ and $b_{s^2}$ contributions are basically orthogonal (see below). The latter should allow for a clear distinction between the standard second order bias picture or its possible extensions discussed above.
\par
As noted above, all of the models under consideration share the same functional form \eqref{eq:densitys2} for the second order halo density field. Only the standard quadratic bias model has $b_{s^2}=0$.  
Therefore, we can write down the bispectrum as
\be
B_\text{mmh}^\text{(unsym)}(\vec k_1,\vec k_2,\vec k_3)-b_1B_\text{mmm}(\vec k_1,\vec k_2,\vec k_3)
=2P(k_1)P(k_2) \biggl[b_2 +b_{s^2} \left(\mu^2-\frac{1}{3}\right) \biggr].
\label{eq:bispectmodel}
\ee
Note that we use a parametrization where the factors of $1/2$ are absorbed into the bias parameters for simplicity. We will restore these prefactors only at the very end, when we are comparing our bias measurements to the theoretical bias functions. A non-vanishing $b_{s^2}$ in the above equation would be a clear evidence for a dynamical biasing picture. After dividing by the two matter power spectra, the remaining quantity is a function of the angle $\mu$ only, which simplifies the combination of information from several scales $k_1$ and $k_2$.
\section{Simulations \& Bispectrum Estimation}
\label{sec:simulations}

\begin{table}[b]
\begin{tabular}{lrrrrrrr}
\hline
\hline
	&$b_1$	&$\Delta b_1$	&$b_2$&$\Delta b_2$	&$b_{s^2}$&$\Delta b_{s^2}$&$M [h^{-1}M_\odot]$\\
\hline
I&$1.142$&$0.002$&$-0.37$&$0.01$&$-0.07$&$0.03$&$9.68\times 10^{12}$\\ 
II&$1.409$&$0.002$&$-0.42$&$0.01$&$-0.21$&$0.04$&$2.90\times 10^{13}$\\ 
III&$1.954$&$0.004$&$-0.12$&$0.02$&$-0.38$&$0.06$&$8.58\times 10^{13}$\\ 
IV&$2.889$&$0.010$&$1.25$&$0.03$&$-0.63$&$0.11$&$2.48\times 10^{14}$\\ 

\hline
\hline
\end{tabular}
\caption{Best fit bias parameters, their errors and mean mass for our four mass bins. The bias parameters are compared to the theoretical bias functions in Figure \ref{fig:massdep}. The first order bias $b_1$ is extracted from the halo-matter cross power spectrum and the second order bias parameters are inferred from the cross bispectrum.
}
\label{tab:bestfit}
\end{table}

\begin{figure}[t]
   \centering
   \includegraphics[width=0.49\textwidth]{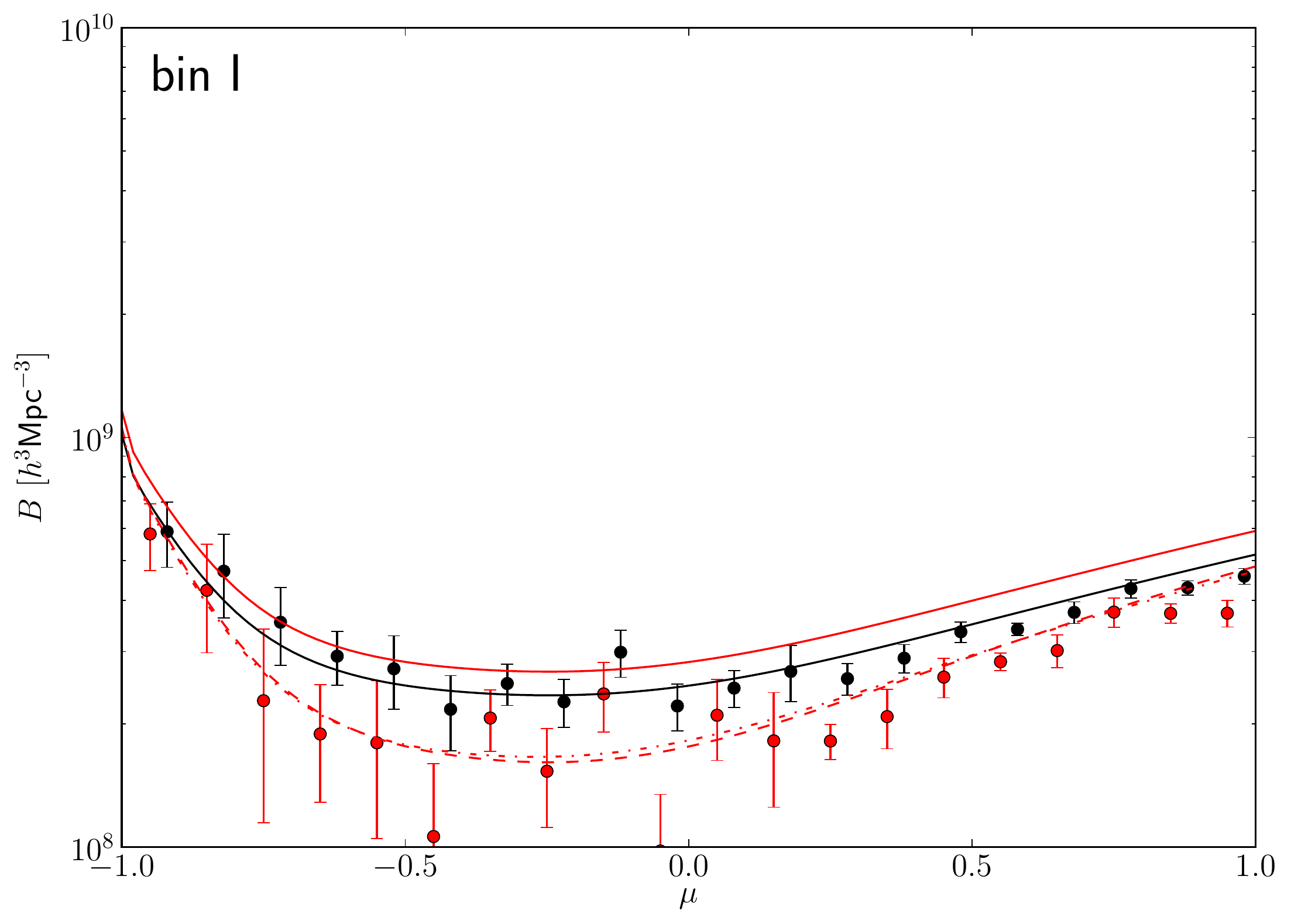} 
   \includegraphics[width=0.49\textwidth]{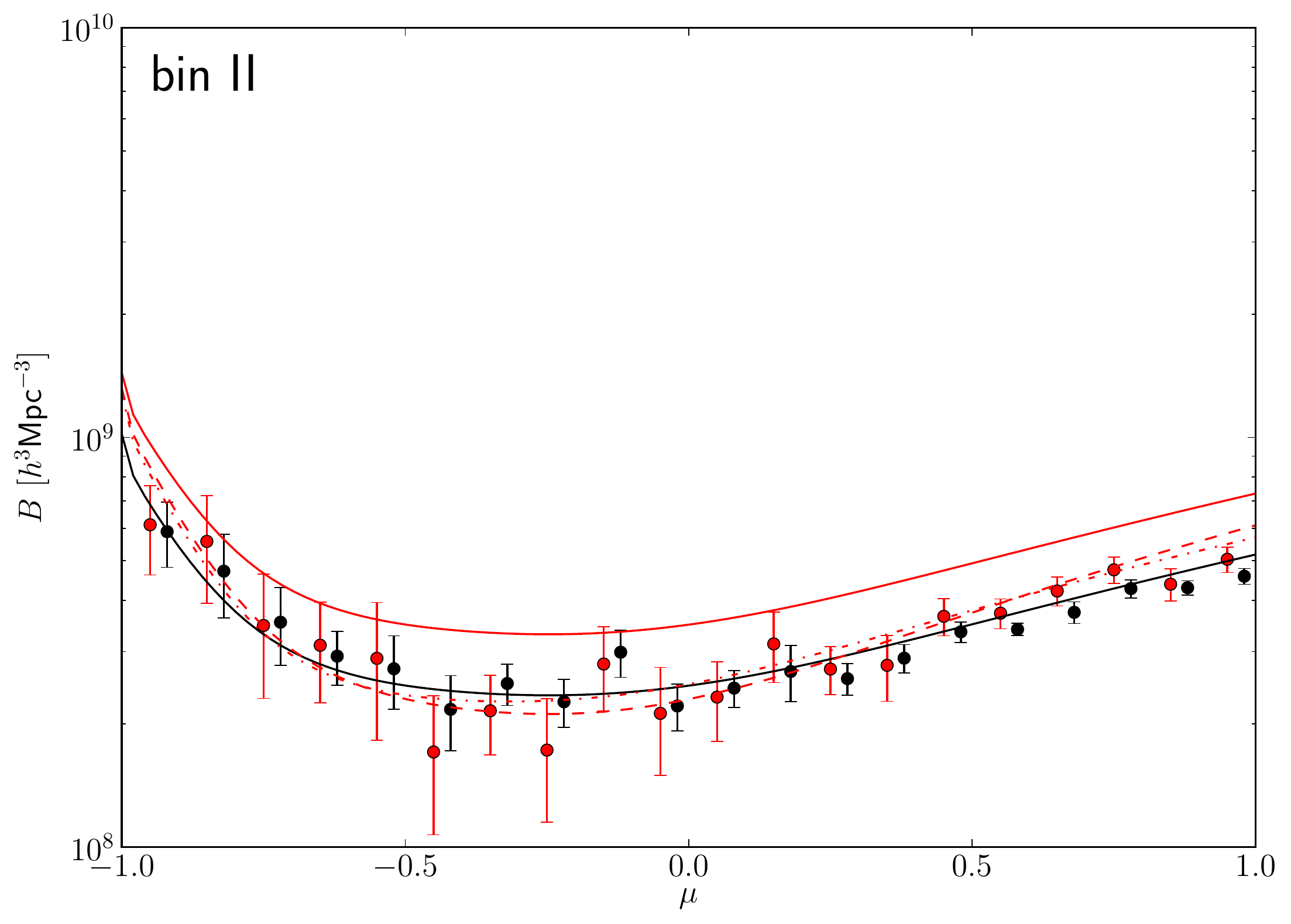} 
   \includegraphics[width=0.49\textwidth]{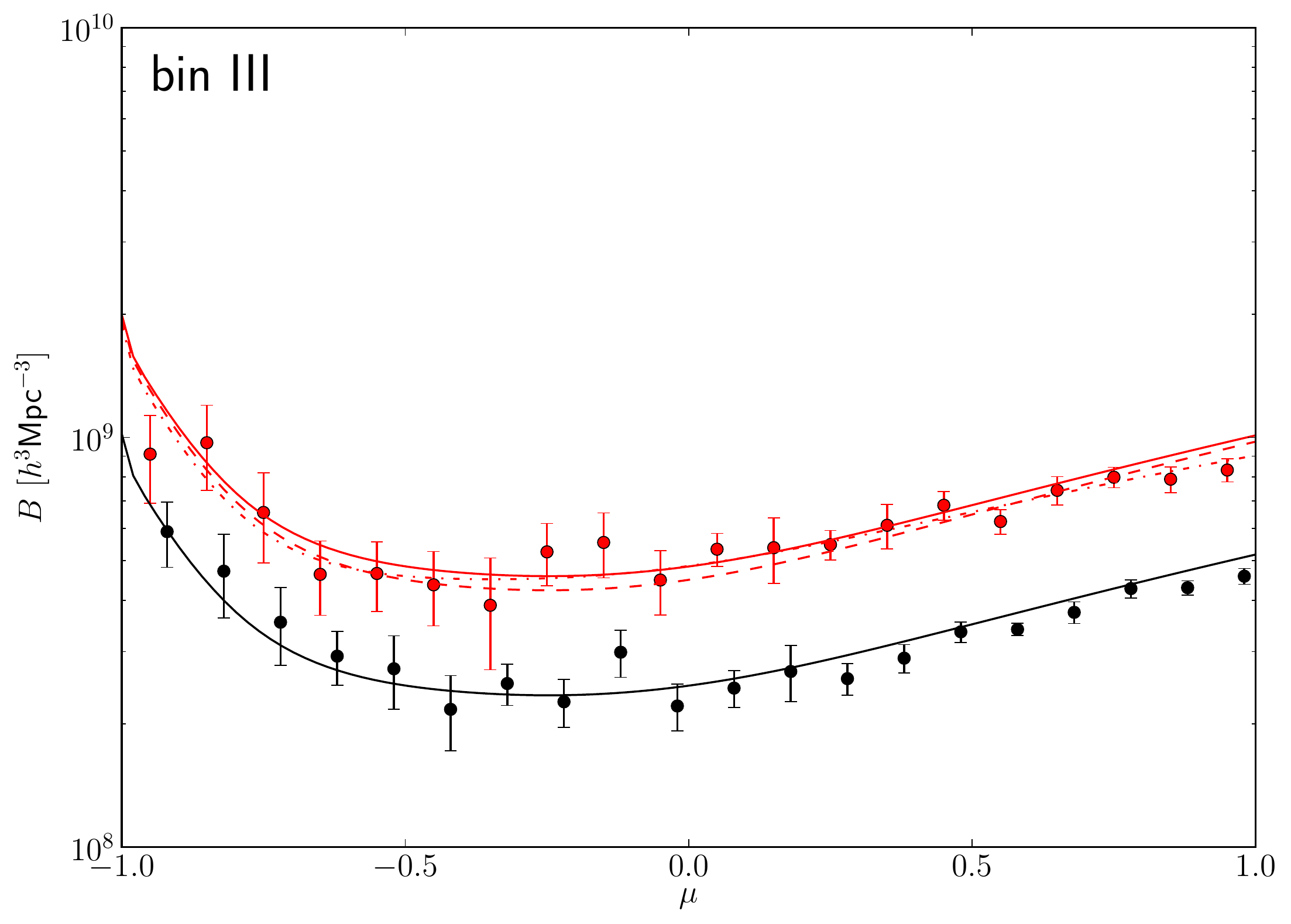} 
   \includegraphics[width=0.49\textwidth]{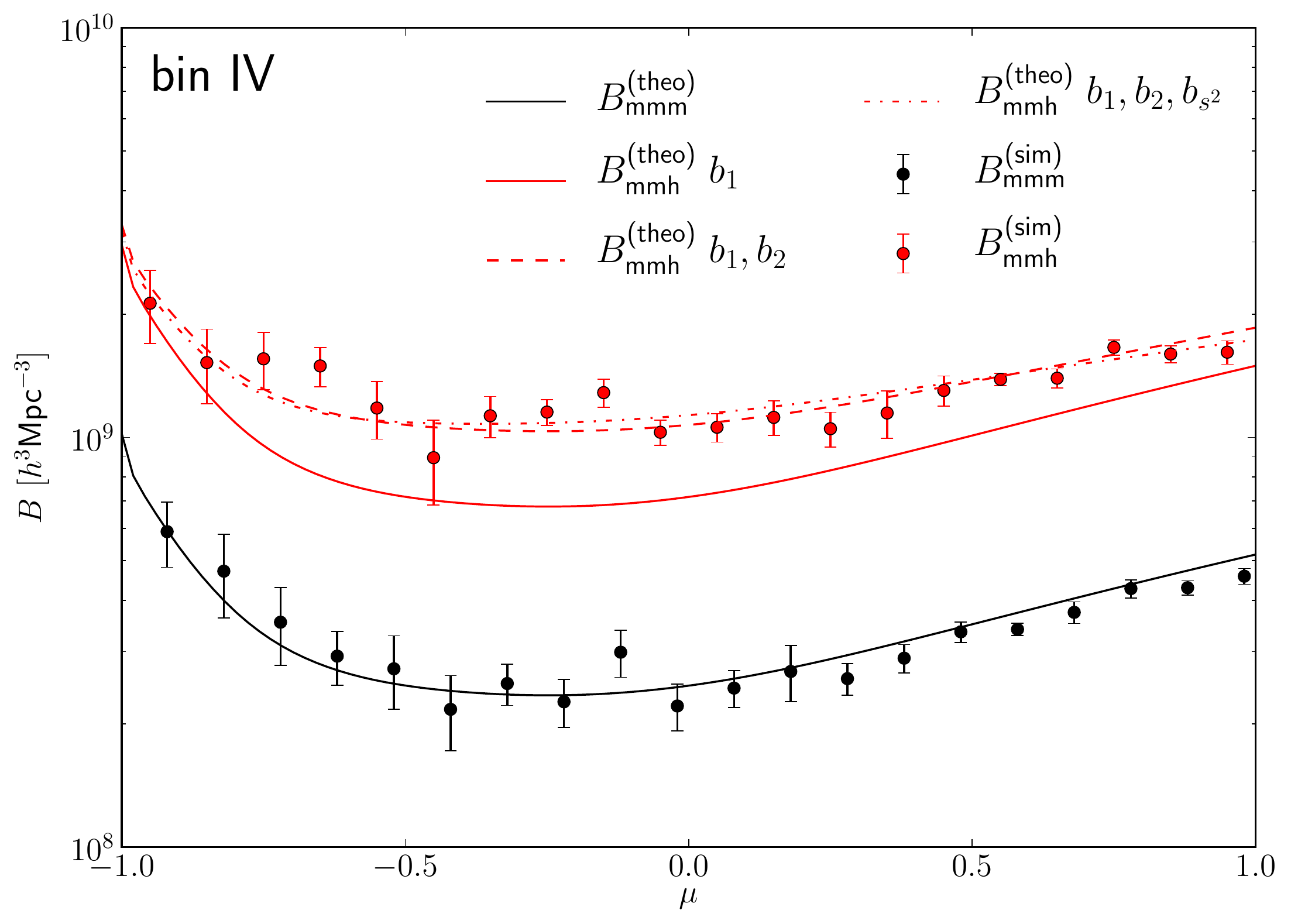} 
   \caption{Matter (black points) and halo-matter-matter (red points) bispectra
as a function of triangle shape for configuration 
$k_1=0.052 \ihMpc$ $k_2=0.06 \ihMpc$. The black solid line shows the tree level
prediction for the matter bispectrum, the red solid line has $b_1$ only and 
dashed and dashed-dotted lines are adding $b_2$ and $b_{s^2}$. The list of bias
parameters behind the theoretical cross bispectra in the legend 
indicates the parameters that were considered for the corresponding curve. The 
error bars are estimated from the box-to-box variance of the bispectrum 
measurement. 
Note that this is only a small fraction of the total bispectrum 
information that our simulations contain, and the log scale also 
de-emphasises what are actually significant differences between the 
fit with and without $s^2$ (these will be highlighted later). 
}
   \label{fig:bispect}
\end{figure}

\begin{figure}[t]
	\centering
	\includegraphics[width=0.49\textwidth]{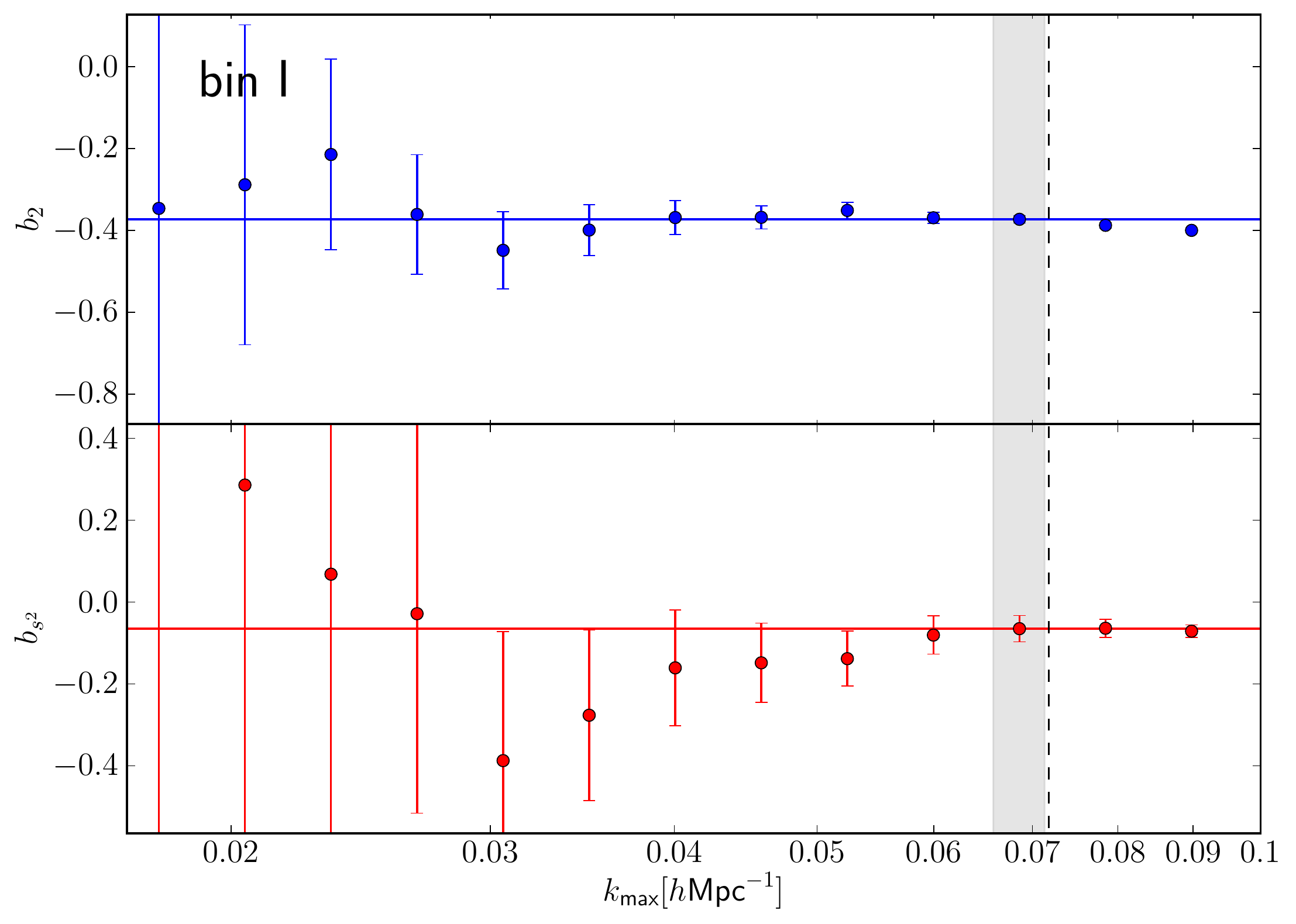} 
	\includegraphics[width=0.49\textwidth]{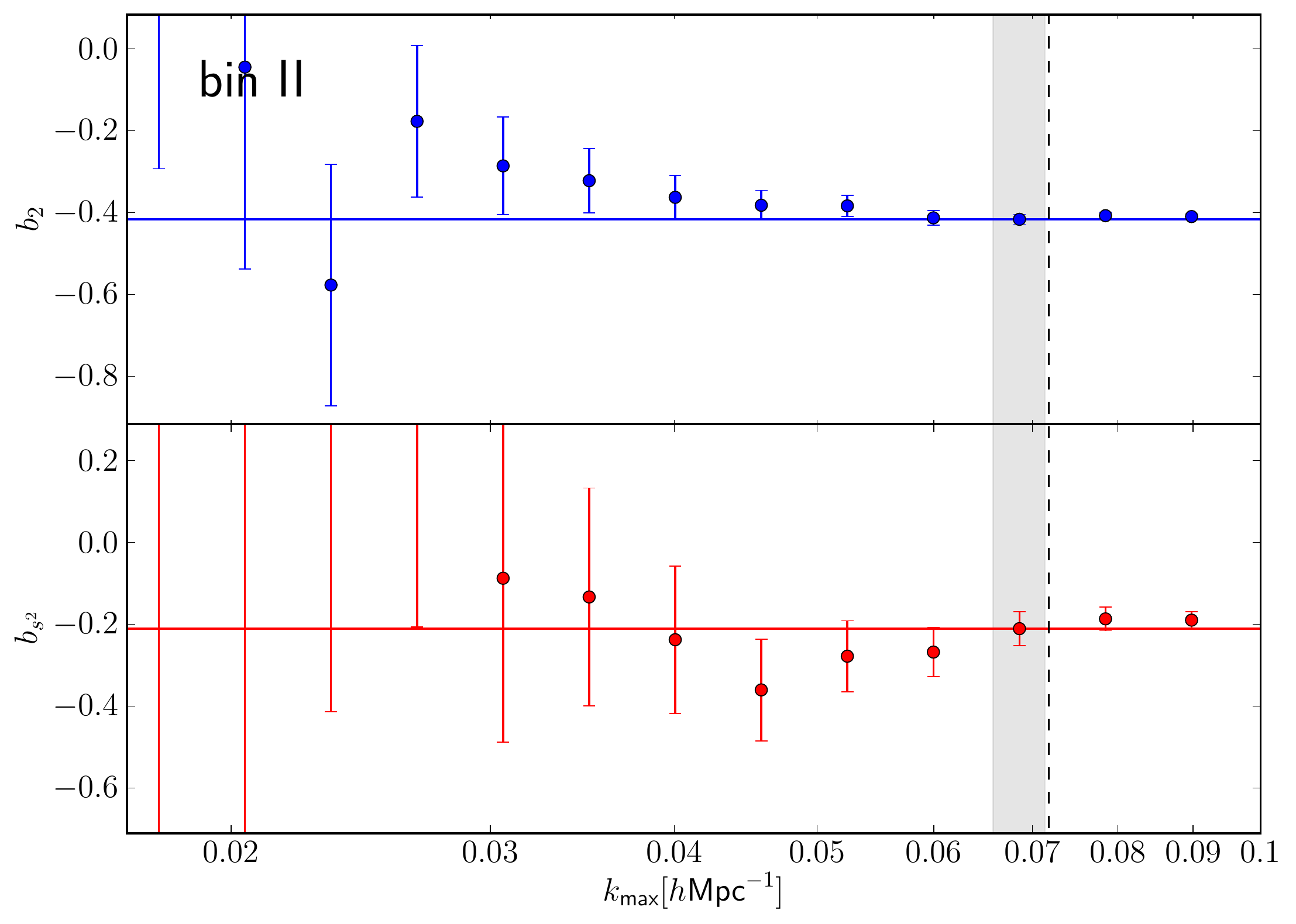}\\
	\includegraphics[width=0.49\textwidth]{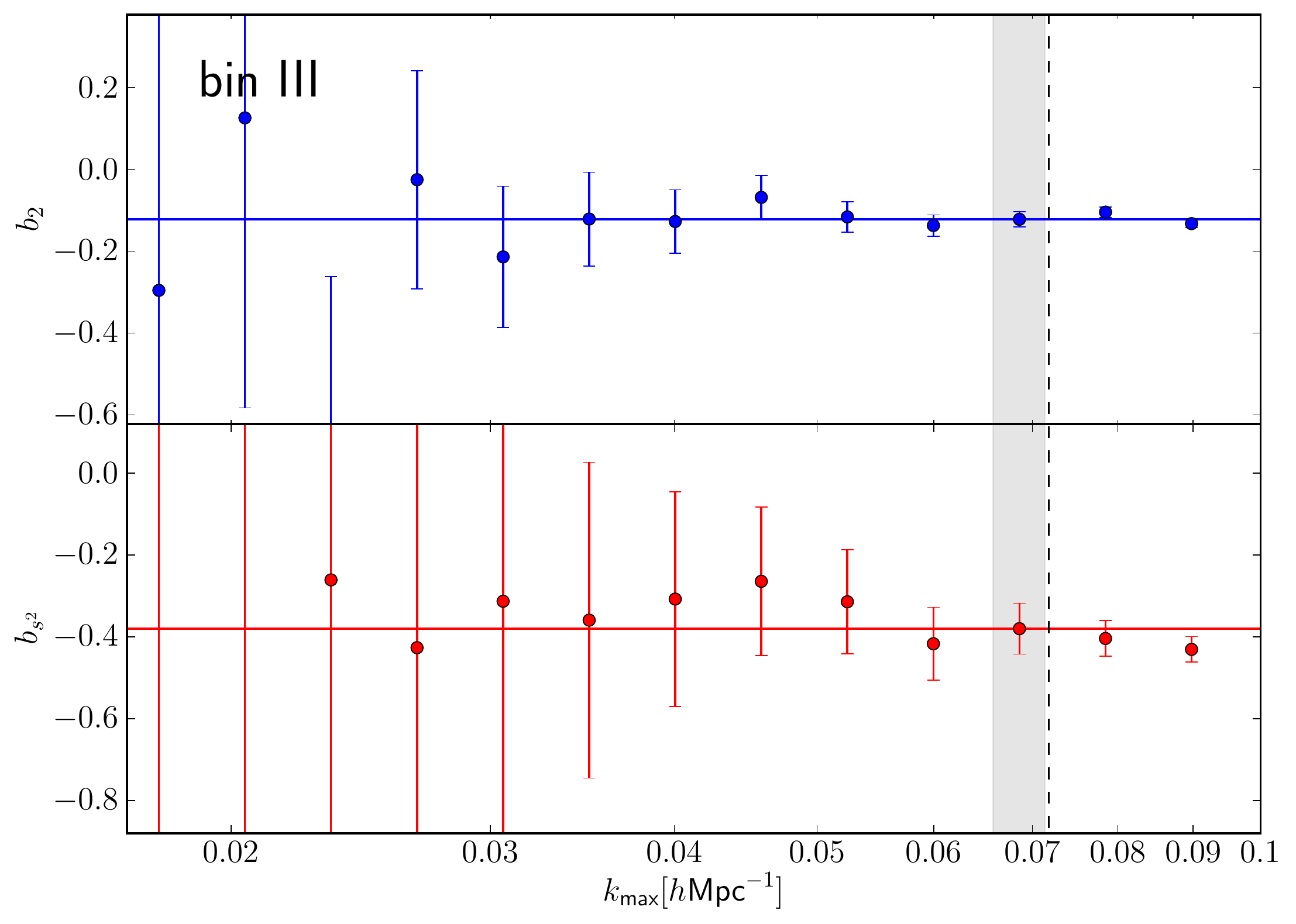} 
	\includegraphics[width=0.49\textwidth]{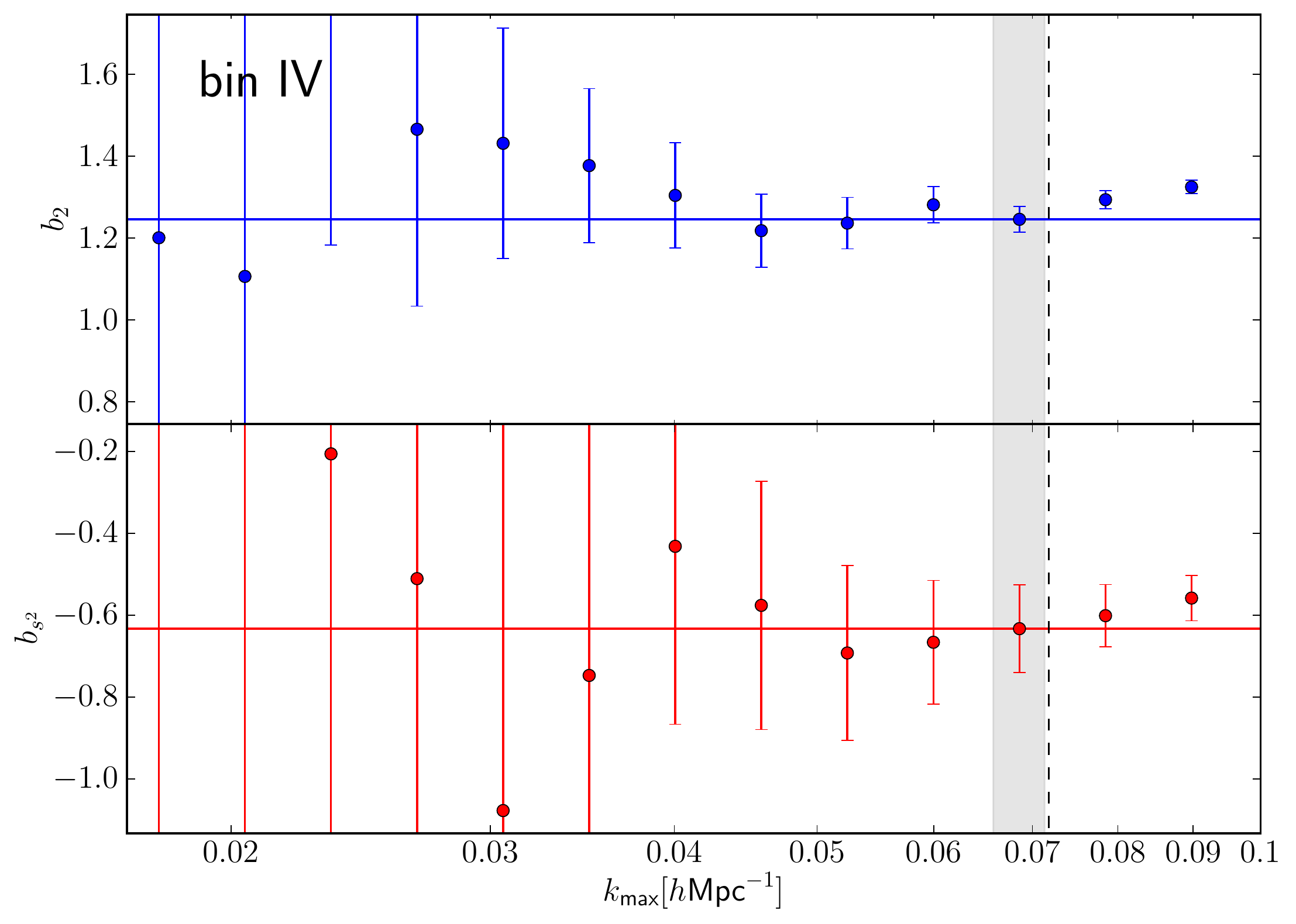} 
	\caption{Convergence of the measured $b_2$ (upper panels) and $b_{s^2}$ (lower panels) parameters with increasing maximum $k$-mode for the four mass bins. The horizontal red and blue lines show the constraints obtained for our fiducial $k_\text{max}=0.07 \ihMpc$. The pivot data point is highlighted by the gray shaded region. }
	\label{fig:converge}
\end{figure}
\begin{figure}[t]
   \centering
   \includegraphics[width=0.49\textwidth]{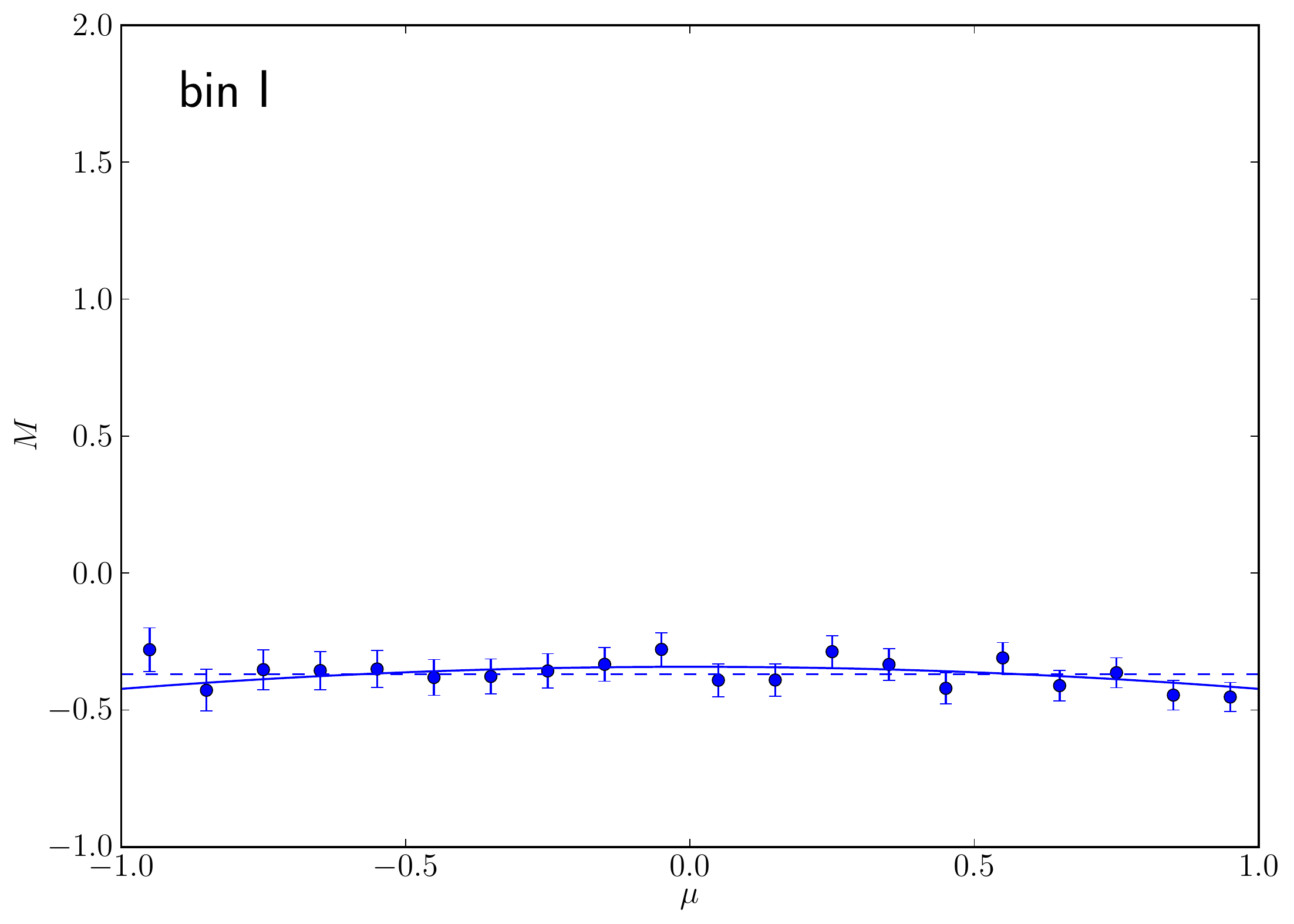} 
   \includegraphics[width=0.49\textwidth]{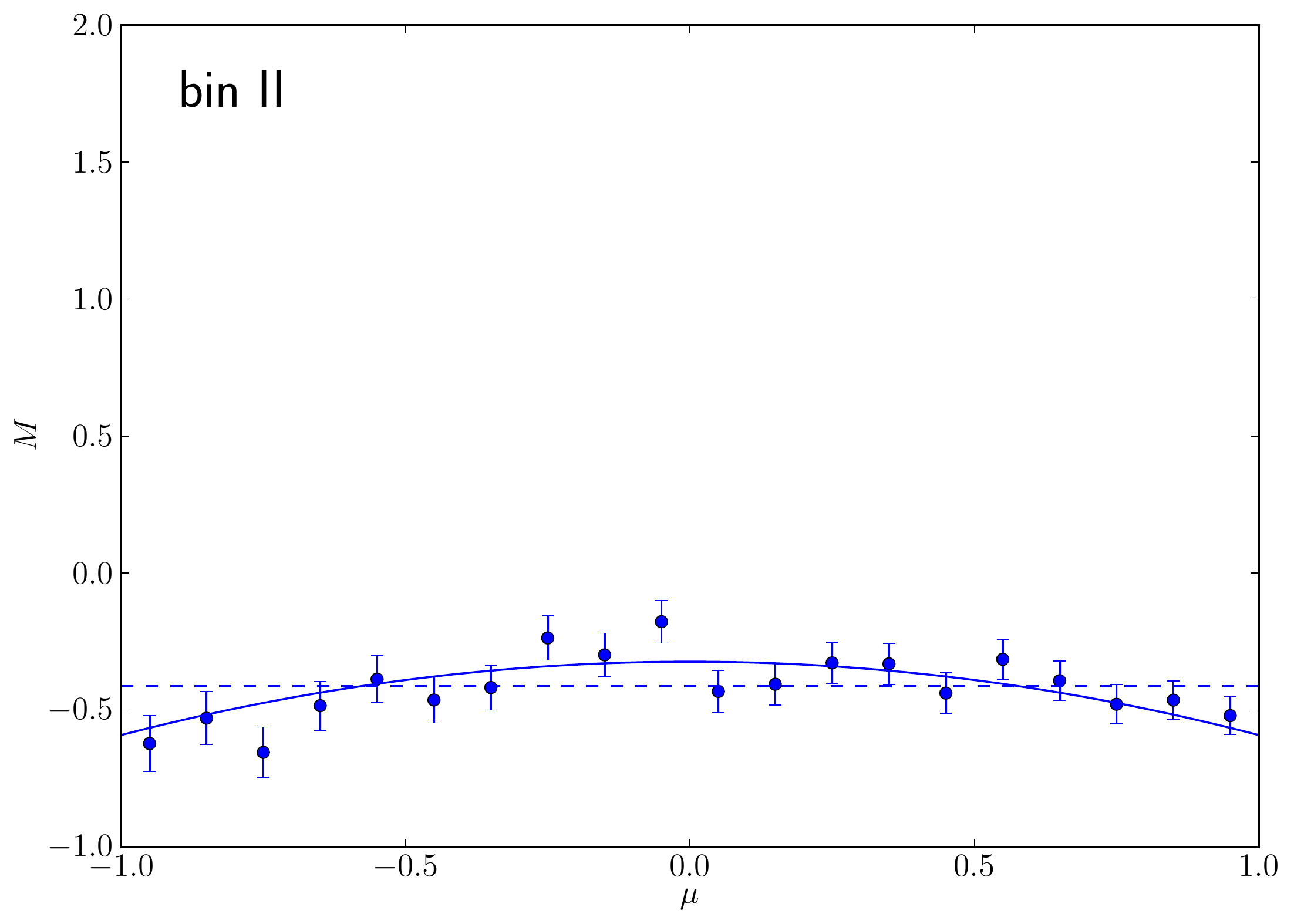} 
   \includegraphics[width=0.49\textwidth]{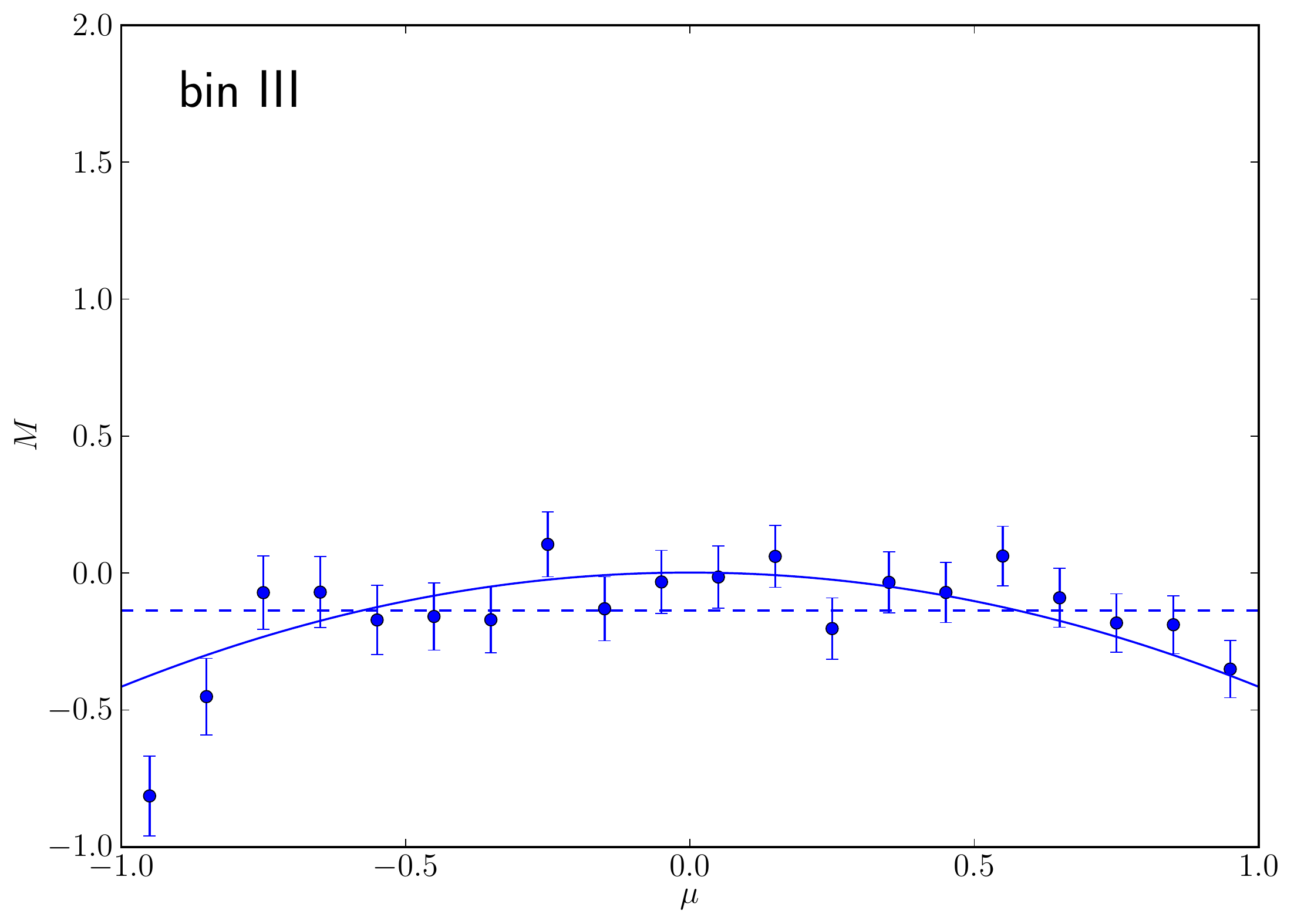} 
   \includegraphics[width=0.49\textwidth]{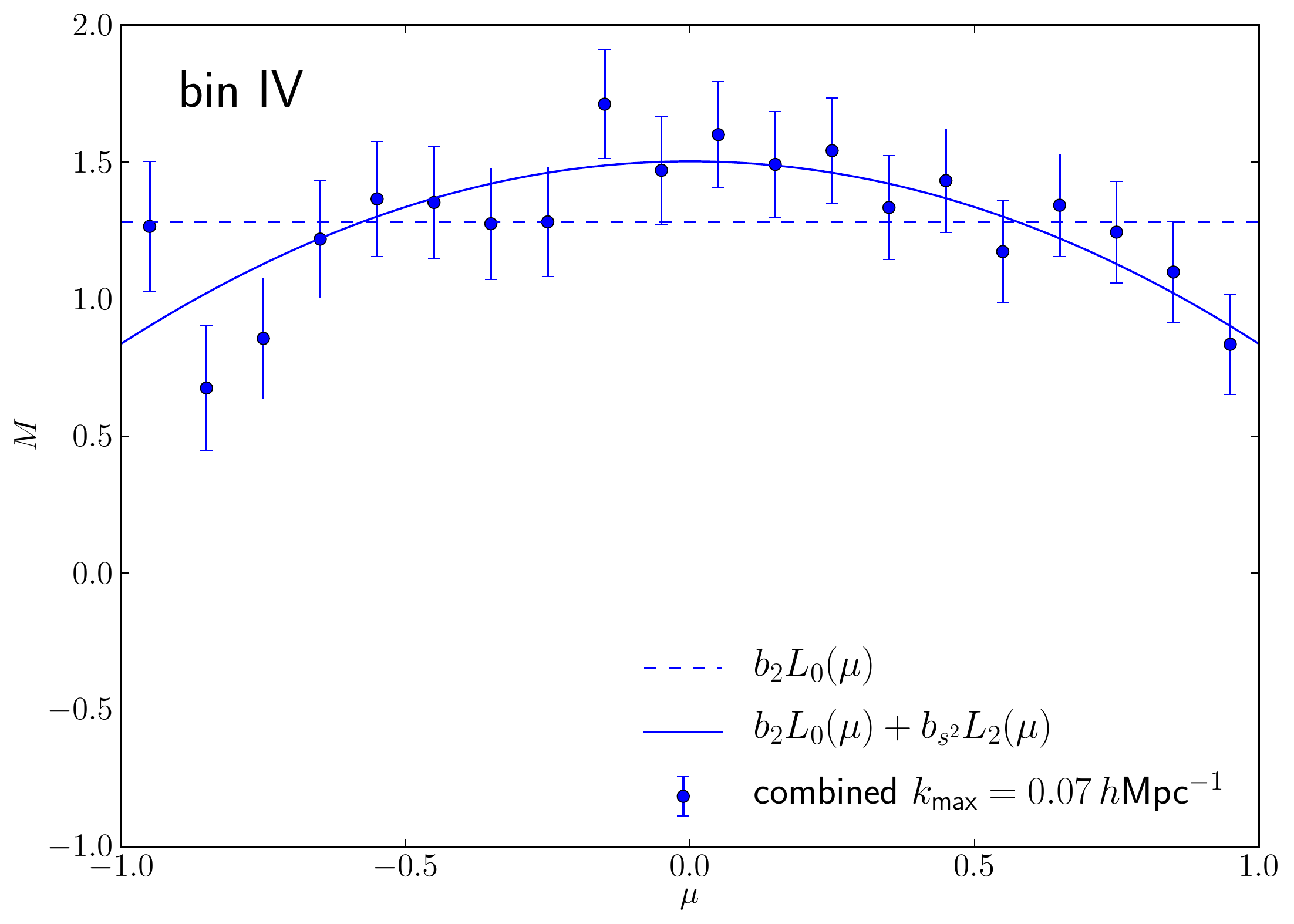} 
   \caption{Residual shape dependence of the halo bispectrum for our reduced 
bispectrum defined in Eq.~\eqref{eq:reducedbispect}. The blue data points with 
error bars show the result of the combined reduced bispectrum defined in 
Eq.~\eqref{eq:muaverage} including all the configurations up to 
$k_\text{max}=0.07 \ihMpc$. The horizontal dashed line shows the model 
including $b_2$ only, the solid blue line shows the model including both $b_2$ 
and $b_{s^2}$. 
}
   \label{fig:resid}
\end{figure}
\begin{figure}[t]
   \centering
      \includegraphics[width=0.49\textwidth]{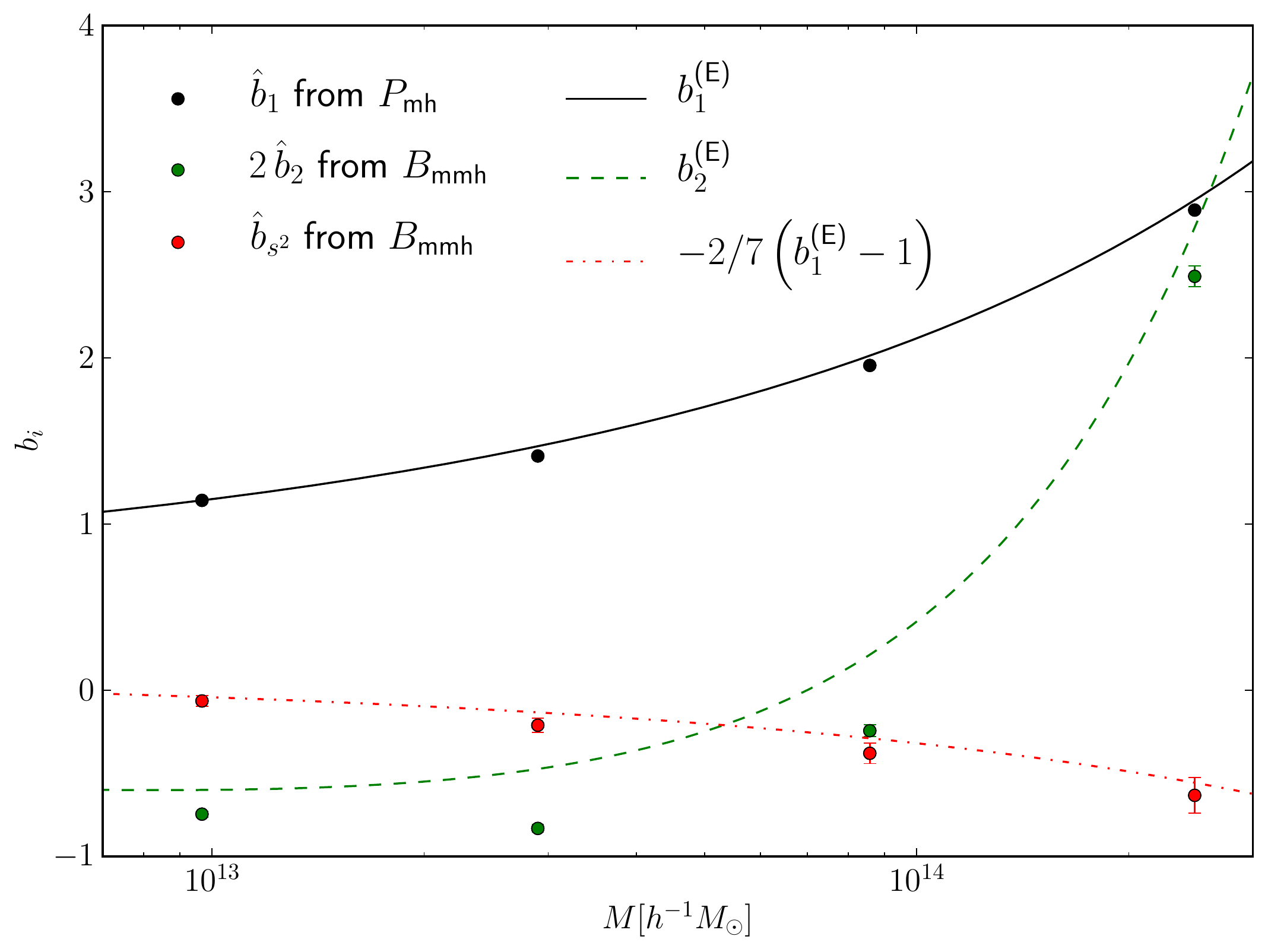} 
      \includegraphics[width=0.5\textwidth]{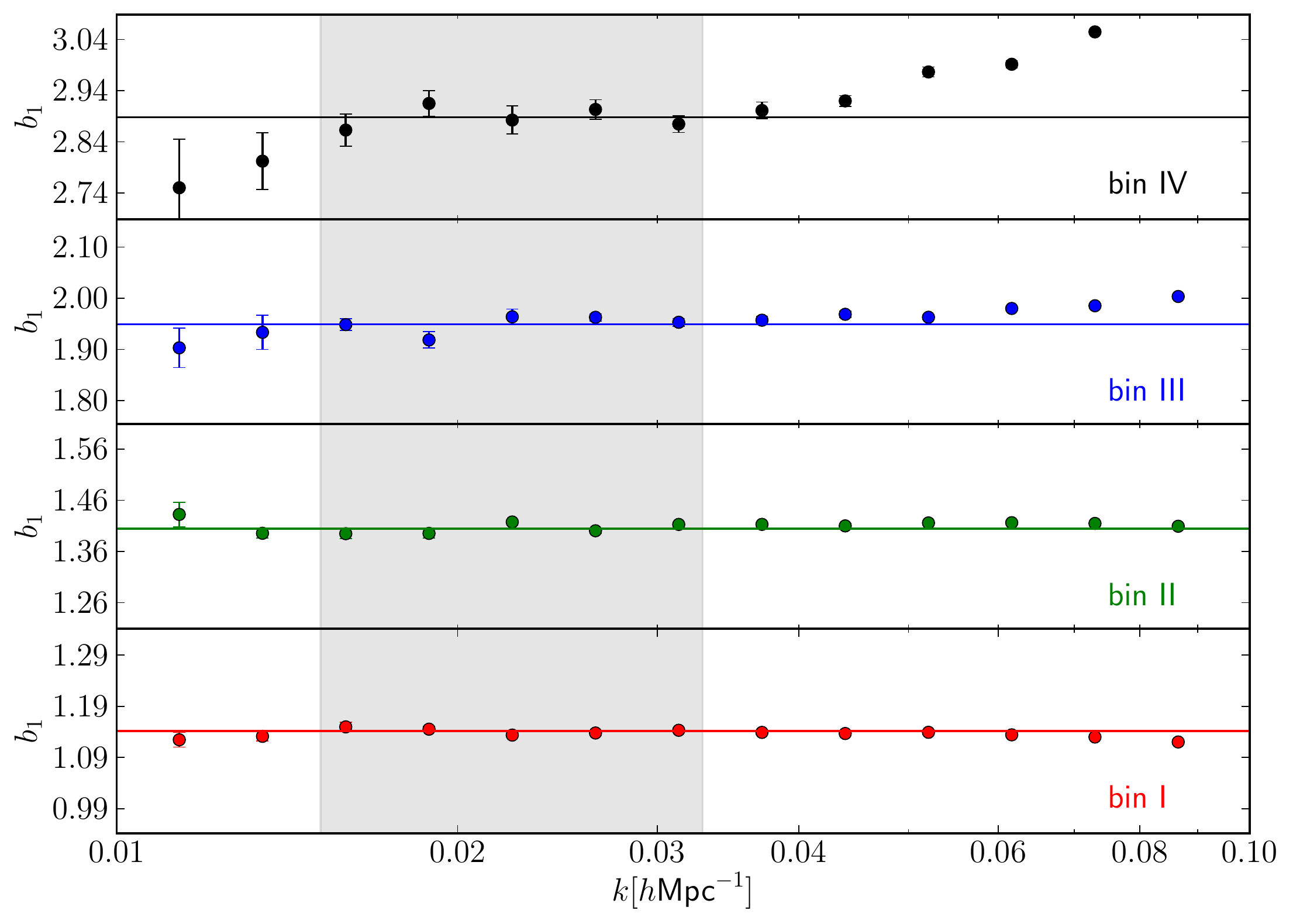}
   \caption{\emph{Left panel: }Mass dependence of the bias parameters and theoretical predictions. The points with error bars are our best fit parameters for $\hat{b}_1$, $2 \hat{b}_2$ and $\hat{b}_{s^2}$. The numerical values of the data points are given in Table \ref{tab:bestfit}. The curves show the corresponding theoretical bias functions as calculated using the relations in \S \ref{sec:biasrelation}. The measurements for $\hat{b}_1$ are in a good agreement with the theory, there is a clear deviation for the $\hat{b}_{s^2}$ and $\hat{b}_2$ measurement for the two central mass bins. \emph{Right panel: }Ratio of the simulation halo matter and matter power spectra $\hat{P}_\text{hm}(k)/\hat{P}_\text{mm}(k)$ and first order bias parameters inferred using the data points highlighted by the shaded region.} 
   \label{fig:massdep}
\end{figure}
\subsection{The Simulations}
We are studying the cosmic density field in a suite of $11$ dark matter only simulations with box size of $L=1600 \hMpc$, which are an extension of the simulations described in \cite{Desjacques:2008vf}. The $\Lambda$CDM model is based on best-fit parameters inferred from the WMAP 5-year data  release \cite{Komatsu:2008hk}. Thus, we adopt a mass density parameter $\Omega_\text{m} = 0.279$, a baryon density parameter $ \Omega_\text{b}= 0.0462$, a Hubble constant $h = 0.7$, a spectral index $n_\text{s} = 0.96$, and a normalization of the curvature perturbations of $\Delta^2_\mathcal{R} = 2.21 \tim{-9}$ at the pivot point $k = 0.02\ \text{Mpc}^{-1}$. This normalization leads to a present day fluctuation amplitude of $\sigma_8 \approx 0.81$. The initial conditions are set up a redshift $z_\text{i}=99$. The gravitational evolution of the $N_\text{p}=1024^3$ particles is integrated using the publicly available \texttt{Gadget2}  code \cite{Springel:2005mi}. Simulations size, particle number and matter density parameter yield a particle mass of $3\tim{11} \hMs$.
Dark matter halos are identified using a Friends-of-Friends halo finder with a linking length of $0.2$ times the mean inter particle spacing. Only halos exceeding $20$ particles are considered for our analysis, corresponding to a minimum halo mass of approximately $6\tim{12} \hMs$. We consider four mass bins, each spanning a factor of three in mass.
For the estimation of the statistics, particles are interpolated on a $N_\text{c}=512$ grid using the Cloud-in-Cell (CIC) algorithm and the gridded density field is corrected for the window of the grid.
The matter-matter-halo and matter bispectrum are measured for low $k$-modes 
$k<0.12 \ihMpc$ in the simulation output at redshift $z=0$. 
The bispectrum measurement scales as the sixth power of the number of grid 
cells per dimension, which makes it computationally very expensive to extract the full bispectrum information at 
higher $k$. It would still be interesting to extent the measurement to higher $k$ in the future to determine the scale of breakdown of the tree level calculation.

\subsection{Bispectrum estimation and data reduction}
The bispectrum modes must satisfy the triangle condition $\vec k_1+\vec k_2+\vec k_3=0$, thus the shape of the bispectrum is fully specified by two lengths and one angle, which we choose as $k_1,k_2$ and $\mu=\vec k_1 \cdot \vec k_2/ k_1 k_2$.
Since $\delta(\vec x)$ is a real valued field, the Fourier modes have to satisfy  $\delta^*(\vec k)=\delta(-\vec k)$. Consequently, the imaginary part of the bispectrum cancels when we add $\delta(\vec k_1)\delta(\vec k_2)\delta(\vec k_3)$ and $\delta(-\vec k_1)\delta(-\vec k_2)\delta(-\vec k_3)$. We consider bins that are logarithmically spaced in $k_1$ and $k_2$ and linearly in $\mu$. We add all the bispectrum amplitudes that fall into the bin centered at $(k_1,k_2,\mu)$.
\par
As a first step in our analysis we subtract out the $b_1$ contribution proportional to the matter bispectrum and divide by the power spectrum measured in the same simulation and $k$-bins to cancel part of the cosmic variance
\be
\hat {M}(k_1,k_2,\mu)=\frac{\hat{B}_\text{mmh}^\text{(unsym)}(k_1,k_2,\mu)-\hat{b}_1 \hat{B}_\text{mmm}(k_1,k_2,\mu)}{2 \hat{P}_\text{mm}(k_1)\hat{P}_\text{mm}(k_2)}.
\label{eq:reducedbispect}
\ee
Comparing to Eq.~\eqref{eq:bispectmodel}, the resulting statistic should be a 
function of $\mu$ only. The hat is used to mark quantities estimated from the 
simulations. This quantity is distinct from the the usual definition of the 
reduced bispectrum since the power spectra in the denominator do not depend on 
$k_3$ and can thus be controlled by limiting $k_1$ and $k_2$. The first order 
bias parameter $\hat{b}_1$ is estimated from the halo-matter cross power 
spectrum on 
large scales $0.015\ihMpc< k <0.03 \ihMpc$, where the linear bias model $P_\text{hm}=b_1 P_\text{mm}$ is believed to be 
accurate. We show the $k$-dependence of $\hat{P}_\text{hm}(k)/\hat{P}_\text{mm}(k)$ and the inferred bias parameters in Fig.~\ref{fig:massdep}. Note that the resulting statistic depends only on the magnitude of the $k_1$ and $k_2$ modes, so that we can ensure the validity of the perturbative expansion by limiting these modes accordingly.

When showing the reduced data as a function of $\mu$ only, we reduce them as 
follows

\be
\chi^2_M=\sum_{k_1,k_2} 
\left(\frac{\hat{M}(k_1,k_2,\mu)-\overline{M}(\mu)}{\Delta M(k_1,k_2,\mu)}
\right)^2
\ee

\be
\overline{M}(\mu)=\sum_{k_1,k_2} \frac{\hat{M}(k_1,k_2,\mu)}{\Delta M^2(k_1,k_2,\mu)}\left(\sum_{k_1,k_2} \frac{1}{\Delta M^2(k_1,k_2,\mu)}\right)^{-1}
\label{eq:muaverage}
\ee
\be
\Delta \overline{M}(\mu)=\left(\sum_{k_1,k_2} \frac{1}{\Delta M^2(k_1,k_2,\mu)} \right)^{-1/2}
\label{eq:deltamuaverage}
\ee
for each $\mu$.
\par
The cosmic variance of the bispectrum estimates could in principle be measured from the standard deviation between our simulation boxes. Given the small number of boxes this approach is bound to give a very noisy estimate. 
Since we are using the error for the weighting of the modes, we would like to avoid a spurious upweighting of modes which by chance have a low simulation to simulation variance. For this purpose we prefer a smooth error estimate.
As shown in \cite{Scoccimarro:2003wn} the variance of the bispectrum is given by
\be
\Delta B_\text{mmh}^2(k_1,k_2,\mu)=s_{123}\frac{V_\text{f}}{V_{123}}\frac{1}{(2\pi)^3}P_\text{mm}(k_1)P_\text{mm}(k_2)\left(P_\text{hh}(k_3)+\frac{1}{\bar n_\text{h}}\right),
\ee
where $V_\text{f}=(2 \pi)^3/L^3$ is the volume of the fundamental cell, $\bar n_\text{h}$ is the number density of the tracer and $s_{123}$ takes on values of 1, 2 and 6 for general, isosceles and equilateral triangles, respectively. The norm volume is given by
\be
V_{123}=\frac{8\pi^2}{(2\pi)^6} k_1^3 k_2^3 \left(\derd \ln k\right)^2 \derd \mu
\ee
for our bins, which are logarithmically spaced in $k_1$, $k_2$ and linearly spaced in $\mu$.
The above just quantifies the diagonals of the covariance matrix between the different triangle shapes and scales, but the correlations between different triangles are believed to be small \cite{Scoccimarro:2003wn}.
When calculating the error on the reduced bispectrum, we focus on the error contribution from the matter-matter-halo bispectrum described above and thus have
\be
\Delta M(k_1,k_2,\mu)\approx\frac{\Delta B_\text{mmh}(k_1,k_2,\mu)}{2P_\text{mm}(k_1)P_\text{mm}(k_2)}
\ee
This procedure could be clearly improved by modeling or measuring the full covariance matrix. 

\subsection{Bias Estimation}
We can now estimate $b_2$ and $b_{s^2}$ minimizing
\be
\chi^2=\sum_{k_1,k_2}\sum_{\mu} \left(\frac{\hat M(k_1,k_2,\mu)-b_2 L_0(\mu)-b_{s^2}L_2(\mu)}{\Delta M(k_1,k_2,\mu)}\right)^2
\ee
where $L_0(\mu)=1$ and $L_2(\mu)=\left(\mu^2-1/3\right)$ are the zeroth and second order Legendre polynomials, which form an orthogonal set on $[-1,1]$. The $k$-sums are performed over $k_\text{min}<k_1<k_\text{max}$, $k_\text{min}<k_2<\kappa k_1$ such that $(1-\kappa) k_1<k_3<(1+\kappa) k_1$ and we use $\kappa =3/4$ for definiteness.

Defining the cosine
\be
\la a(k_1,k_2,\mu),b(k_1,k_2,\mu)\ra:=\sum_{k_1,k_2}\sum_\mu \frac{a(k_1,k_2,\mu) b(k_1,k_2,\mu)}{\Delta M^2(k_1,k_2,\mu)}
\ee
we obtain for the best fit parameters
\begin{align}
\hat{b}_2=&\frac{\la L_2,L_2\ra\langle L_0, \hat{M}\rangle}{\Delta}-\frac{\la L_0,L_2\ra\langle L_2,\hat{M}\rangle}{\Delta}\\
\hat{b}_{s^2}=&-\frac{\la L_0,L_2\ra\langle L_0, \hat{M}\rangle}{\Delta}+\frac{\la L_0,L_0\ra\langle L_2, \hat{M}\rangle}{\Delta},
\end{align}
where $\Delta=\la L_0, L_0\ra\la L_2, L_2\ra-\la L_0 ,L_2 \ra^2 $. Note that from $\chi^2$ minimization one 
obtains an equivalent expression for the cosine of the reduced data
\be
\la a,b\ra_\mu:=\sum_\mu \frac{a(\mu)b(\mu)}{\Delta \overline{M}^2(\mu)},
\ee
where for $a(\mu),b(\mu)$ we have $\la a(\mu),M(k_1,k_2,\mu)\ra=\la a(\mu),\overline{M}(\mu)\ra_\mu$ and $\la a(\mu),b(\mu)\ra=\la a(\mu),b(\mu)\ra_\mu$.

The bias parameter $b_1$ is estimated from the halo-matter cross power spectrum on large scales/low k, 
where loop corrections are believed to be unimportant. 
Furthermore there is no shotnoise contamination in the cross power spectrum. 
We use $b_1$ measured from the halo-matter cross-power spectrum and use it for the 
bispectrum calculation,
because it has a small statistical error.
There is a correlation between the $b_1$ used for the matter bispectrum subtraction and the inferred $b_2$ and $b_{s^2}$ parameters. The matter bispectrum and the residual considered for our reduced bispectrum should in principle not be degenerate, since the matter bispectrum has additional $L_1(\mu)$ terms and since the coefficients of the three Legendre polynomials in the matter bispectrum depend on the ratio $k_2/k_1$. However, applying our fitting method to the matter bispectrum we obtain $b_2$ and $b_{s^2}$ constraints of order unity. Thus an error of $\Delta b_1$ in the first order bias leads to a shift of roughly $-\Delta b_1$ for both $b_2$ and $b_{s^2}$. The statistical errors on the second order bias coefficients are roughly a factor of five larger than the statistical error on the corresponding first order bias, such that this effect is not very important from the statistical point of view. 
We would only need to worry about it if the first order bias coefficients in the power spectrum and bispectrum differ.
While this may at first glance seem impossible (and it probably is), 
the higher order loop terms renormalize $b_1$ \cite{McDonald:2006mx} and 
to our knowledge it has not been explicitly shown that the renormalization works the same way in bispectrum as in 
power spectrum. In this paper we will assume that the two are the same. 

The errors on the estimated parameters are calculated from the expected deviation in the $\chi^2$. For a $n$-component parameter vector $\vec a$ the $\chi^2$ in the vicinity of the best fit parameter set $\vec{\hat{a}}$ can be described as
\be
\chi^2(\vec a)=\chi^2(\vec{\hat{a}})+\frac{1}{2}\sum_{i,j} (a_i-\hat{a}_i)\frac{\partial^2 \chi^2}{\partial a_i \partial a_j}(a_j-\hat{a}_j),
\ee
where $\Delta \chi^2=\left|\chi^2(\vec a)-\chi^2(\vec{\hat{a}})\right|=1$ corresponds to one sigma errors on the parameters. Thus
\begin{align}
\Delta b_2=\sqrt{1/\la L_0,L_0\ra} && \Delta b_{s^2}=\sqrt{1/\la L_2,L_2\ra}
\end{align}
We compared these errors with the standard deviation of the bias constraints obtained from the single realizations and found good agreement.
\section{Results}\label{sec:results}
Figure \ref{fig:bispect} shows the matter bispectrum and matter-matter-halo bispectra for our four mass bins for one configuration of $k_1=0.052 \ihMpc$, $k_2=0.06\ihMpc$ as a function of the opening angle $\mu$.
The matter bispectrum is in quite good agreement with the theoretical prediction in Eq.~\eqref{eq:bmmmtheo}. There is some small tension for positive $\mu$, but this should not be a problem for our study since our bias constraints are not relying on this theoretical modeling since we are subtracting out the matter bispectrum measured from the simulations.

We also plot the matter-matter-halo cross bispectrum and the theoretical model of Eq.~\eqref{eq:densitys2}. To visualize the effect of the contributions of the bias parameters, we plot the model with vanishing $b_2$ and $b_{s^2}$, with vanishing $b_{s^2}$ and the full model. The non-vanishing bias parameters for the theory lines were chosen according to our best fit model discussed below. While the $k$-values were chosen quite high to reduce the errors, there is still too much scatter in the data points to decide whether the model with and without $b_{s^2}$ gives a better description of the data. This motivated the careful combination of all the information available by weighting the modes accordingly, as described above.

Our main reason to study lowest order non-linear biasing in the bispectrum was 
that these terms are the leading order terms on large scales. As one includes 
higher momenta, loop corrections gain importance and need to be modeled 
accordingly. To asses the importance of higher order corrections and to show 
the convergence of our fitting procedure, we perform our parameter estimation 
as a function of the maximum wavenumber and show the results in 
Figure \ref{fig:converge}. The error bars clearly shrink as we go to higher $k$
 and the inferred bias parameters are almost always consistent with our fiducial 
result shown by the horizontal lines. In these plots we also show the scale up 
to which we have a complete measurement of all the modes by the vertical dashed
line and highlight the $k_\text{max}$ that we use for the primary reported 
parameter values by the vertical shaded region.

Figure \ref{fig:resid} shows our reduced bispectrum $\overline{M}(\mu)$ defined in Eq.~\eqref{eq:muaverage} as a function of opening angle $\mu$ and the combined errors according to Eq.~\eqref{eq:deltamuaverage}. We overplot the $b_2$ only model and the model including both $b_2$ and $b_{s^2}$. These plots show clear evidence for the presence of the tidal term except for the lowest mass bin, for which the $b_2$ only model gives an acceptable description of the angular dependence.

In Table \ref{tab:bestfit} we give the best fit values of the first and second order bias parameters for our four halo mass bins obtained considering all the modes up to $k_\text{max}=0.07 \ihMpc$. Figure \ref{fig:massdep} shows the bias parameters as a function of mass together with the corresponding predictions of the spherical collapse model. Note that we are plotting $2 \hat{b}_2$, which corresponds the second order Eulerian bias (see Eq.~\eqref{eq:quadraticbias}). 
The $\hat{b}_1$ measurements from the halo-matter cross power spectrum are in good agreement with the theoretical predictions. 
The measured $\hat{b}_{s^2}$ are slightly lower than the theoretical predictions for the two central mass bins, but the trend with mass is well reproduced by the theory. The $\hat{b}_2$ measurements are less well reproduced by the theoretical bias function, especially mass bins II and III are well below the theory. The theoretical predictions for $b_1$ and $b_{s^2}$ are given by first derivatives of the mass function and do not reproduce the data perfectly. Thus one would naturally expect some corrections for the second derivatives. The disagreement could also be an indication for a failure of the spherical collapse picture at second order.
Specifically, the fact that the predicted $b_{s^2}$ disagrees with the measurements suggests that Langrangian $b_{s^2}^\text{(L)}$
is not zero, as predicted by the 
ellipsoidal collapse model \cite{1979ApJ...231....1W,1996ApJS..103....1B,ShethMoTormen,Desjacques:2007zg}. 

The right panel of Fig.~\ref{fig:massdep} shows the ratio of halo-matter and matter-matter power spectra used for the inference of the first order bias. We are fitting for $\hat{b}_1$ on large scales to avoid the regime where the non-linear corrections become important. These corrections are affecting the highest mass bin quite strongly already 
starting at $k\approx 0.05 \ihMpc$. The corrections are stronger for higher 
mass objects, in accordance with the general mass dependence of the second order bias parameters derived here, but a full 
discussion of all the terms entering at one loop level must also include the third order terms, which is beyond the scope of this paper. 

As in the power spectrum analysis, the bispectrum is increasingly affected by loop corrections as one increases the maximum momentum in the problem. The relevant quantity here is the largest external momentum involved and thus one should make sure that all the $k$'s are in the perturbative regime and can be well described by the order of perturbation theory considered. Our study focuses on the large scale bispectrum, where a tree level treatment should be sufficiently accurate and we have shown that the constraints are both stable and consistent as we increase $k_\text{max}$ up to $0.1\ihMpc$.

\section{Discussion \& Outlook}\label{sec:discussion}
In this paper we presented a dynamical motivation for the second order tidal
tensor bias term proposed in \cite{McDonald:2009dh}, showing that it is naturally 
generated by gravity even if absent in the initial conditions.
We performed a measurement of this bias for dark matter halos in 
simulations, showing clear evidence for the tidal tensor bias, 
increasing with the halo mass. 
Our results are consistent with the 
picture 
in which a significant part of the tidal tensor bias is 
generated by the gravitational evolution, but we also find some 
evidence that it is present already in the initial conditions. 
While the functional form of the additional terms 
agrees with the discussion in \cite{McDonald:2009dh} the dynamical derivation 
can supplement it by a prediction of the time dependence and gives at least 
qualitative 
understanding about the mass dependence of the bias parameters. 

Our analysis also includes second order density bias $b_2$, which we find to disagree with the theoretical predictions 
at a quantitative level, even if it qualitatively follows the predictions that it should be negative at low mass and 
increase with halo mass. 
The disagreement calls for a reinvestigation and improvement of the theoretical bias function or measurement of the second order bias parameters in the initial conditions, where the Lagrangian bias parameters are postulated to describe the density field of the protohaloes. 
The deviation between the theoretical and measured $b_2$ parameters does not necessarily mean that the Lagrangian picture is wrong, but rather that the bias functions derived from the spherical collapse model and the mass function might not be sufficiently accurate.

Because these two terms are quadratic, they contribute  a loop correction to the power spectrum, so the standard linear bias picture still applies on very large scales, even if they contribute at the leading order to the bispectrum.
If we want to understand halo power spectrum on smaller scales then the loop corrections caused by these quadratic 
terms become important. 
However, a consistent calculation of the halo correlation function or power spectrum at one loop level 
requires a model of the halo density field up to third order, including all the non-local terms \cite{McDonald:2009dh}. 
This gives rise to several other terms that contribute to the power spectrum 
at the same order, each with a prefactor that may depend on the halo mass. 
A clean extraction of these non-linear bias coefficients from the power spectrum  alone
is very difficult and instead higher order correlations are needed to separate 
these terms. 
This paper is a first step in extracting the quadratic coupling terms from the 
bispectrum. The next step in this program is to extract the cubic order terms from the trispectrum: such an analysis will be presented elsewhere. 
For this reason a discussion of the implications on the two point function based on our 
results for the quadratic couplings cannot be complete. Still, there are some obvious implications of our results. 
One is that the non-local tidal tensor bias contributions must be included in the analysis of galaxy power spectrum 
and ignoring them may lead to incorrect conclusions. This effect is specially relevant for the broadband power, but 
will also have an effect on the position of the baryonic acoustic oscillations (BAO), 
just like the quadratic density term does if $b_2 \ne 0$ \cite{PadWhite}. Our preliminary calculations suggest this effect is subdominant, with about a factor of 4 lower prefactor in front of $b_{s^2}$ relative to the $b_2$ term, suggesting it will not strongly affect the 
measurement of the true BAO scale.  

\begin{acknowledgments}
As this paper was being completed the draft by \cite{Chan:2011} appeared, some 
of which is based on a similar analysis with some similar conclusions. 
The authors would like to thank Teppei Okumura, Ravi Sheth, Zvonimir Vlah and Matias Zaldarriaga for discussions. TB would like to thank the Berkeley Center for Cosmological Physics and the Lawrence Berkeley Laboratory for the kind hospitality. VD acknowledges support from the Swiss National Science Foundation. This work is supported by the DOE, the Swiss National Foundation under contract 200021-116696/1 and WCU grant R32-10130. The simulations were performed on the ZBOX3 supercomputer of the Institute for Theoretical Physics at the University of Zurich.
\end{acknowledgments}

\appendix

\section{Matter density in LPT up to second order}\label{app:lpt}
In the above derivation of the final halo density field arising from a local bias in Lagrangian space we used the Lagrangian displacement field and the second order matter density in SPT. In \cite{Matsubara:2007wj} it is shown that after expanding the exponential damping prefactor, the one loop power spectra in LPT and SPT are identical. This appendix shows that this equivalence is true also in terms of the fields \emph{if} one expands the LPT expressions up to the desired order in the density field. Without these expansions, LPT was shown to contain a non-trivial resummation of SPT terms.
\par
For $n$-th order displacement field one has \cite{Matsubara:2007wj}
\be
{}^{(n)}\vec \Psi(\vec k,\eta)=-\frac{\ii}{n!} \int \frac{\derd^3 p_1}{(2\pi)^3}\ldots \int \frac{\derd^3 p_n}{(2\pi)^3} {}^{(n)}\vec L(\vec p_1,\ldots,\vec p_n)\ \delta(\vec p_1,\eta)\ldots \delta(\vec p_n,\eta)\ (2\pi)^3\delta^\text{(D)}\left(\vec k-\sum \vec p_i\right),
\ee
with the kernels
\begin{align}
{}^{(1)}\vec L(\vec p)=\frac{\vec p}{p^2}, && {}^{(2)}\vec L(\vec p_1,\vec p_2)= \frac{3}{7}\frac{\vec p_1+\vec p_2}{|\vec p_1+\vec p_2|^2}\left[1-\left(\frac{\vec p_1 \cdot \vec p_2}{p_1 p_2}\right)^2\right].
\end{align}
The density field in $k$-space can be obtained upon Fourier transforming the Eulerian field in configuration space and using the Jacobian mapping $\bigl[1+\delta(\vec x,\eta)\bigr]\derd^3 x=\derd^3 q$
\be
\delta(\vec k,\eta)=\int \derd^3 q \eh{\ii \vec k\cdot \vec q}\left\{\eh{\ii \vec k \cdot \vec \Psi(\vec q,\eta)}-1\right\}
\ee
We can now expand the exponential up to second order in the displacement field
\begin{align}
\delta(\vec k,\eta)\approx& \int \derd^3 q \eh{\ii \vec k\cdot \vec q}\left\{\ii \vec k \cdot \vec \Psi(\vec q,\eta)-\frac{1}{2}(\vec k \cdot \vec \Psi(\vec q,\eta))^2\right\}\\
=&\ii \vec k \cdot {}^{(1)}\vec \Psi(\vec k,\eta)+\ii \vec k \cdot {}^{(2)}\vec \Psi(\vec k,\eta)+\frac{1}{2} \int \frac{\derd^3 k'}{(2\pi)^3}\frac{\vec k \cdot \vec k'}{(\vec k')^2}\frac{\vec k \cdot (\vec k-\vec k')}{|\vec k-\vec k'|^2}{}^{(1)}\delta(\vec k',\eta){}^{(1)}\delta(\vec k -\vec k',\eta)\\
=& {}^{(1)}\delta(\vec k,\eta)+\int \frac{\derd^3 k'}{(2\pi)^3} F_2(\vec k',\vec k-\vec k') {}^{(1)}\delta(\vec k',\eta){}^{(1)}\delta(\vec k -\vec k',\eta)
\end{align}
Thus, expanding the exponential we see that LPT and SPT agree at second order in the density field.

\bibliography{s2bispect}
\end{document}